\documentstyle[eqsecnum,prd,aps,epsf]{revtex}

\draft
\begin{document}

\newcommand{\beq}{\begin{equation}}
\newcommand{\eeq}{\end{equation}}
\newcommand{\beqn}{\begin{eqnarray}}
\newcommand{\eeqn}{\end{eqnarray}}
\newcommand{\pa}{\partial}
\newcommand{\vp}{\varphi}
\newcommand{\varep}{\varepsilon}
\newcommand{\ep}{\epsilon}

\twocolumn[\hsize\textwidth\columnwidth\hsize\csname
@twocolumnfalse\endcsname

\title{Stability and collapse of rapidly rotating,
supramassive neutron stars: \\3D simulations in general relativity}

\author{Masaru Shibata$^{1,2}$, Thomas W. Baumgarte$^1$, and 
Stuart L. Shapiro$^{1,3}$}

\address{$^1$ Department of Physics, 
University of Illinois at Urbana-Champaign, Urbana, IL 61801}

\address{$^2$ Department of Earth and Space Science,~Graduate School of
Science,~Osaka University, Toyonaka, Osaka 560-0043, Japan}

\address{$^3$ Department of Astronomy and NCSA, 
University of Illinois at Urbana-Champaign, Urbana, IL 61801}

\maketitle

\begin{abstract}
We perform 3D numerical simulations in full general relativity to
study the stability of rapidly rotating, supramassive neutron stars at
the mass-shedding limit to dynamical collapse.  We adopt an adiabatic
equation of state with $\Gamma = 2$ and focus on uniformly rotating
stars.  We find that the onset of dynamical instability along
mass-shedding sequences nearly coincides with the onset of secular
instability.  Unstable stars collapse to rotating black holes within
about one rotation period.  We also study the collapse of stable stars
which have been destabilized by pressure depletion (e.g.~via a phase
transition) or mass accretion.  In no case do we find evidence for the
formation of massive disks or any ejecta around the newly formed Kerr
black holes, even though the progenitors are rapidly rotating.
\end{abstract}

\vskip2pc]

\baselineskip 6.2mm

\section{Introduction}

Neutron stars found in nature are rotating.  Rotation can support
stars with higher mass than the maximum static limit, producing
``supramassive'' stars, as defined and calculated by Cook, Shapiro and
Teukolsky~\cite{CST2,CST4}.  Such supramassive stars can be created
when neutron stars accrete gas from a normal binary companion.  This
scenario can also lead to ``recycled'' pulsars [see~\cite{CST3} for
model calculations in general relativity].  Alternatively,
supramassive stars can be produced in the merger of binary neutron
stars (\cite{BSS} for discussions and references). 

Pulsars are believed to be uniformly rotating.  Eventually, viscosity
will drive any equilibrium star to uniform rotation.  Uniformly
rotating configurations with sufficient angular momentum will be
driven to the mass-shedding limit (at which the star's equator rotates
with the Kepler frequency, so that any further spin-up would disrupt
the star;~\cite{CST3}).  Supramassive neutron stars at the
mass-shedding limit is the subject of this paper.

The dynamical stability of rotating neutron stars, including
supramassive configurations, against radial perturbations, as well as
the final fate of unstable stars undergoing collapse, has not been
established definitively.

Along a sequence of nonrotating, spherical stars, parameterized by
central density, the maximum mass configuration defines a critical
density above which the stars are unstable to radial oscillations:
stars on the high density, unstable branch collapse to black holes on
dynamical timescales~\cite{chandra,MTW,ST}.

Establishing the onset of instability for rotating stars is more
complicated.  Chandrasekhar and Friedman~\cite{CF} and
Schutz~\cite{Schutz} developed a formalism to identify points of {\em
dynamical} instability to axisymmetric perturbations along sequences
of rotating stars.  In their formalism, however, a complicated
functional for a set of trial functions has to be evaluated. Probably
because of the complexity of their methods, explicit calculations have
never been performed.  Friedman, Ipser and Sorkin~\cite{FIS} showed
that for uniformly rotating stars the onset of {\em secular}
instability can be located quite easily by applying turning-point
methods along sequences of constant angular momentum.  This method has
been applied to find points of secular instability in numerical models
of uniformly rotating neutron stars~\cite{CST2,CST4,Ster}.

Turning point methods along such sequences can only identify the point
of secular, and not dynamical instability, since one is comparing
neighboring, uniformly rotating configurations with the same angular
momentum.  Maintaining uniform rotation during perturbations tacitly
assumes high viscosity.  In reality, the star will preserve
circulation as well as angular momentum in a dynamical perturbation,
and not uniform rotation.  It is thus possible that a secularly
unstable star may be dynamically stable: for sufficiently small
viscosity, the star may change to a differentially rotating, stable
configuration instead of collapsing to a black hole.  Ultimately, the
presence of viscosity will bring the star back into rigid rotation,
driving the star to an unstable state.  A secular instability evolves
on a dissipative, viscous timescale, while a dynamical instability
evolves on a collapse (free-fall) timescale.  Friedman, Ipser and
Sorkin~\cite{FIS} showed that along a sequence of uniformly rotating
stars, a secular instability always occurs {\em before} a dynamical
instability (implying that all secularly stable stars are also
dynamically stable).

For spherical stars, the onset of secular and dynamical instability
coincides (since for a nonrotating star a radial perturbation
conserves both circulation and uniform rotation).  This suggests that
for uniformly rotating stars for which the rotational kinetic energy
$T$ is typically a small fraction of the gravitational energy $W$, the
onset of dynamical instability is close to the onset of secular
instability.  One goal of this paper is to test this hypothesis and
to identify the onset of dynamical instability in rotating stars.

We also follow the nonlinear growth of the radial instability and
determine the final fate of unstable configurations.  Nonrotating
neutron stars collapse to black holes, but rotating stars could also
form black holes surrounded by massive disks.  Also, if $J/M_g^2$
exceeds the Kerr limit of unity (where $J$ is the angular momentum and
$M_g$ the total mass-energy or gravitational mass of the progenitor
star), not all of the matter can collapse directly to a black hole.
As pointed out recently, such a system could be the central source of
$\gamma$-ray bursts~\cite{vietri}.

Numerical hydrodynamic simulations in full general relativity (GR)
provide the best approach to understanding the collapse of
rotating neutron stars.  Two groups~\cite{Nakamura,SP} included
rotation in axisymmetric, relativistic hydrodynamic codes to study
the collapse of rotating massive stars to black holes.
The collapse and fate of unstable rotating neutron stars, however, has
never been simulated before.  Probably this is because numerical
methods for constructing initial data describing rapidly rotating
neutron stars, as well as numerical tools, techniques and sufficient
computational resources have only become available
recently.  Over the last few years, robust numerical techniques for
constructing equilibrium models of rotating neutron stars in full GR
have been established~\cite{CST2,CST4,Ster,FIP,KEH,Eric}.  More recently,
methods for the numerical evolution of 3D gravitational fields have
been developed (see, e.g.,~\cite{ON,SN,Getal,Cetal,BS2,FMST,gw3p2}).
In a previous paper~\cite{gr3d}, Shibata presented a wide variety of
numerical results of test problems for his 3D hydrodynamic GR code and
demonstrated that simulations for many interesting problems are now
feasible.

In this paper, we perform simulations in full GR for rapidly rotating
neutron stars.  This study is a by-product of our long-term effort to
build robust, fully relativistic, hydrodynamic evolution codes in 3D.
We adopt rigidly rotating supramassive neutron stars at mass-shedding
as initial data.  By exploring rotating stars at mass-shedding, we can
clarify the effect of rotation most efficiently.  Such stars are also
the plausible outcome of pulsar recycling and binary coalescence.
Following Ref.~\cite{gr3d}, we prepare equilibrium states for such
stars using an approximation in which the spatial metric is assumed to
be conformally flat.  We then perform numerical simulations to
investigate the dynamical stability of the rapidly rotating neutron
stars against collapse and to determine the final fate of the unstable
neutron stars.  We believe that this is the first 3D simulation of the
dynamical collapse of a rotating neutron star in full GR.

In Newtonian physics, stars with sufficient rotation ($T/|W| \gtrsim
0.27$) are dynamically unstable to bar
formation~\cite{Chandra2,LRS,CT}.  Since $T/|W|$ increases
approximately with $R^{-1}$ as a star collapses, radial collapse may
drive the dynamical growth of nonaxisymmetric bars.  To allow for this
possibility, a numerical simulation must be performed in 3D, not in
axisymmetry.

In Sec. II, we briefly describe our formulation, initial data, and
spatial gauge conditions.  In Sec. III, we present numerical results.
First, we study the dynamical stability of supramassive rotating
neutron stars at the mass-shedding limit.  We then study the final
products of the unstable neutron stars adopting three kinds of initial
conditions: In the first case, we choose a marginally stable neutron
star and slightly reduce the pressure for destabilization as the
initial condition.  In the second case, we prepare a stable star, and
then reduce a large fraction of the pressure suddenly.  While we are
primarily interested in computational consequences, this scenario may
provide a model for sudden phase transitions inside neutron
stars~(see, e.g.,~\cite{Bethe,Glendenning} and references therein).
In the third case, we prepare a stable star and add more mass near the
surface to induce collapse.  In all the cases, we find that the final
products are black holes without surrounding massive disks, which we
can readily explain.  In Sec. IV we provide a summary.  
Throughout
this paper, we adopt the units $G=c=M_{\odot}=1$ where $G$, $c$ and
$M_{\odot}$ denote the gravitational constant, speed of light and
solar mass, respectively.  We use Cartesian coordinates $x^k=(x, y,
z)$ as the spatial coordinate with $r=\sqrt{x^2+y^2+z^2}$; $t$ denotes
coordinate time.

\section{Methods}

\subsection{Summary of formulation}

We perform numerical simulations using the same formulation as
in~\cite{gr3d}, to which the reader may refer for details and basic
equations.  The fundamental variables used in this paper are:
\beqn
\rho &&:{\rm rest~ mass~ density},\nonumber \\
\varep &&: {\rm specific~ internal~ energy}, \nonumber \\
P &&:{\rm pressure}, \nonumber \\
u^{\mu} &&: {\rm four~ velocity}, \nonumber \\
\alpha &&: {\rm lapse~ function}, \nonumber \\
\beta^k &&: {\rm shift~ vector}, \nonumber \\
\gamma_{ij} &&:{\rm metric~ in~ 3D~ spatial~ hypersurface},\nonumber \\ 
\gamma &&=e^{12\phi}={\rm det}(\gamma_{ij}), \nonumber \\
\tilde \gamma_{ij}&&=e^{-4\phi}\gamma_{ij}, \nonumber \\
K_{ij} &&:{\rm extrinsic~curvature}.\nonumber 
\eeqn
General relativistic hydrodynamic equations are solved using 
the van Leer scheme for the advection terms \cite{Leer}. 
Geometric variables (together with three auxiliary functions $F_i$ and the
trace of the extrinsic curvature) are evolved with a free evolution
code. The boundary conditions for geometric variables are 
the same as those adopted in \cite{gr3d}. 
Violations of the Hamiltonian constraint and conservation of
mass and angular momentum are monitored as code checks.  Several test
calculations, including spherical collapse of dust, stability
of spherical neutron stars, and the evolution of rotating neutron
stars as well as corotating binary systems have been described
in~\cite{gr3d}.  Black holes that form during the late phase of the
collapse are located with an apparent horizon finder as described
in~\cite{S}. 

We also define a density $\rho_*(=\rho \alpha u^0 e^{6\phi})$ from
which the total rest mass of the system can be integrated as
\beq
M_*=\int d^3 x \rho_*. 
\eeq
We perform the simulations assuming $\pi$-rotation
symmetry around the $z$-axis as well as a reflection symmetry with
respect to the $z=0$ plane and using a fixed uniform grid with the typical
size $153 \times 77 \times 77$ in $x-y-z$. We have also performed 
test simulations with different grid resolutions to check that the 
results do not change significantly. 

The slicing and spatial gauge conditions we use in this paper are
basically the same as those adopted in~\cite{gr3d}; i.e., we
impose an ``approximate'' maximal slice condition ($K_k^{~k} \simeq 0$)
and an ``approximate'' minimum distortion gauge condition ($\tilde D_i
(\pa_t \tilde \gamma^{ij}) \simeq 0$ where $\tilde D_i$ is the
covariant derivative with respect to $\tilde \gamma_{ij}$). However,
for the case when a rotating star collapses to a black hole, we
slightly modify the spatial gauge condition in order to improve the
spatial resolution around the black hole.
The method of the modification is described in Sec. II.C.

\subsection{Initial conditions for rotating neutron stars}

As initial conditions, we adopt rapidly and rigidly rotating
supramassive neutron stars in (approximately) equilibrium states.  The
approximate equilibrium states are obtained by choosing a conformally
flat spatial metric, i.e., assuming $\gamma_{ij}=e^{4\phi}
\delta_{ij}$.  This approach is computationally convenient 
and, as illustrated in~\cite{CST}, provides an excellent approximation
to exact axisymmetric equilibrium configurations.

Throughout this paper, we assume a $\Gamma$-law equation of state in
the form
\beq
P=(\Gamma-1)\rho \varep,
\eeq 
where $\Gamma$ is the adiabatic constant. For hydrostatic problems, 
the equation of state can be rewritten in the polytropic form
\beq
P = K \rho^{\Gamma}, \mbox{~~~~~} \Gamma = 1 + \frac{1}{n} \label{eos},
\eeq
where $K$ is the polytropic constant and $n$ the polytropic index.  We
adopt $\Gamma=2$ ($n = 1$) as a reasonable qualitative approximation
to realistic (moderately stiff) cold, nuclear equations of state.

Physical units enter the problem only through the polytropic constant
$K$, which can be chosen arbitrarily or else completely scaled out of
the problem.  We often quote values for $K = 200/\pi$, for which in
our units ($G = c = M_{\odot} = 1$) the radius is $R = (\pi K/2)^{1/2}
= 10$ in the Newtonian limit; corresponding to $R \sim 15$ km.  Since
$K^{n/2}$ has units of length, dimensionless variables can be
constructed as
\beqn \label{rescale}
\bar M_* = M_* K^{-n/2}, & \bar M_g =  M_g K^{-n/2}, & 
\bar R = R K^{-n/2}, \nonumber  \\
\bar J = J K^{-n},  & \bar {\rm P} = {\rm P} K^{-n/2}, &
\bar{\rho} = \rho K^n,
\eeqn
where ${\rm P}$ denotes rotational period.  All results can be scaled for
arbitrary $K$ using Eqs.~(\ref{rescale}).

For the construction of the (approximate) equilibrium states as
initial data, we adopt a grid in which the semi major axes of the
stars, along the $x$ and $y$-axes, are covered with 40 grid points.
For rotating stars at mass-shedding near the maximum mass, the
semi minor (rotation) axis along the $z$-axis is covered with $23$ or
24 grid points.

In Fig.~1, we show the relation between the gravitational mass $M_g$
and the central density $\rho_c$ for the neutron stars.  The solid and
dotted lines denote the relations for spherical neutron stars and
stars rotating at the mass-shedding limit, constructed from the exact
stationary matter and field equations.  The open circles denote the
approximate equilibrium states at the mass-shedding limit obtained using
the conformal flatness approximation.  We find that at $\rho_c =
\rho_{\rm max}$ where $0.0040 \alt \rho_{\rm max} \alt 0.0045$, $M_g$
takes its maximum value.  For stars with stiff equations of state the
numerical results in Ref.~\cite{CST4} show that 
the central density at the onset of secular
instability (which hereafter we refer to as $\rho_{\rm crit}$)  
is very close to $\rho_{\rm max}$ (see, e.g., Fig.~4 
of~\cite{CST4}).  We therefore consider
stars with $\rho_c \geq \rho_{\rm max} (\simeq \rho_{\rm crit})$ 
as candidates for dynamically unstable stars.

\subsection{Spatial gauge condition}

When no black hole is formed, we adopt the approximate minimum
distortion gauge condition as our spatial gauge condition (henceforth
referred to as the AMD gauge condition). However, as pointed out in
previous papers \cite{gw3p2,gr3d}, during the black hole formation
(i.e., for an infalling radial velocity field), the expansion of the
shift vector $\pa_i \beta^i$ and the time derivative of $\phi$ becomes
positive using this gauge condition together with maximal
slicing. Accordingly, the coordinates diverge outward and the
resolution around the black hole forming region becomes worse and
worse during the collapse.  This undesirable property motivates us to
modify the AMD gauge condition when we treat black hole formation.
Specifically, we modify the AMD shift vector according to
\beq
\beta^i=\beta^i_{\rm AMD}-f(t,r){x^i \over r+\epsilon}
\beta^{r'}_{\rm AMD}.
\eeq
Here $\beta^i_{\rm AMD}$ denotes the shift vector 
in the AMD gauge condition, $\beta^{r'}_{\rm AMD}
\equiv x^k\beta^k_{\rm AMD}/(r+\epsilon)$, $\epsilon$ is a 
small constant much less than the grid spacing, 
and $f(t,r)$ is a function chosen as
\beq
f(t,r)=f_0(t){1 \over 1+(r/M_{g,0})^4}.
\eeq 
where $M_{g,0}$ denotes the gravitational mass of a system 
at $t=0$.  
We determine $f_0(t)$ from $\phi_0=\phi(r=0)$. 
Taking into account the fact that the resolution around $r=0$ 
deteriorates when $\phi_0$ becomes large, we choose $f_0$ according to
\beq
f_0(t)=\left\{
\begin{array}{ll}
\displaystyle 
1 & {\rm for}~\phi_0 \geq 0.8,\\
2.5\phi_0 -1& {\rm for}~0.4 \leq \phi_0 \leq 0.8, \mbox{~~Type I} \\
0 & {\rm for}~\phi_0 < 0.4, \\
\end{array}
\right. \label{type1}
\eeq
or 
\beq
f_0(t)=\left\{
\begin{array}{ll}
\displaystyle 
1 & {\rm for}~\phi_0 \geq 0.6,\\
5\phi_0 -2& {\rm for}~0.4 \leq \phi_0 \leq 0.6, \mbox{~~Type II} \\
0 & {\rm for}~\phi_0 < 0.4. \\
\end{array}
\right.\label{type2}
\eeq
Note that for spherical collapse with $f_0 = 1$, $\pa_i \beta^i \sim
0$ at $r = 0$ in both cases.  In general, we find numerically
that $\pa_i \beta^i$ is small near the origin, where the collapse
proceeds nearly spherically.  In the following, we refer to the
modified gauge conditions of $f_0$ defined by Eqs.~(\ref{type1}) and
(\ref{type2}) as type I and II, respectively. We employ them whenever
a rotating neutron star collapses to a black hole.

\section{Numerical results}

\subsection{Dynamical stability}

We investigate the stability of supramassive rotating neutron stars at
mass-shedding limits against gravitational collapse.  We adopt the
stars marked with (A), (B), (C), (D), and (E) in Fig. 1 as initial
conditions for our numerical experiments.  The physical properties of
these stars are summarized in Table I.  In this numerical experiment,
we adopt two initial conditions for each model. In one case, we use
the (approximate) equilibrium states of rotating neutron stars without
any perturbation and in the other case, we uniformly reduce the
pressure by $1\%$ (by decreasing $K$; i.e., $\Delta K/K=1\%$ where
$\Delta K$ denotes the depletion factor of $K$~\cite{footnote}). 

In Fig. 2, we show $\rho$ and $\alpha$ at $r=0$ 
\cite{footrho} as a function of
$t/{\rm P}$ where ${\rm P}$ is the rotation period of each rotating
star.  We find that when $\rho_c < \rho_{\rm crit}$ (i.e., stars (A),
(B) and (C)), the rotating stars oscillate independent of the
initial perturbations. Hence, these stars are stable against
gravitational collapse.  We note that we find small amplitude
oscillations even when we do not reduce the pressure initially.  This
is caused by small deviations of the initial data from true
equilibrium states, both because of the conformal flatness
approximation and because of numerical truncation error.

We expect the oscillation frequencies in Fig. 2 to be the fundamental
quasi-radial oscillation of these rotating stars.  The oscillation
periods increase with the central density.  At the marginally stable
point of secular stability ($\rho_c=\rho_{\rm crit}$), the period
becomes infinite.

Star (D) does not collapse either and instead oscillates for
$\Delta K=0$.  However, it collapses for $\Delta K/K=1\%$.  This
indicates that star (D) is located near the onset point of
dynamical stability. It is found that the oscillation amplitude for
the case $\Delta K=0$ is very large compared with those for
(A)--(C). This could be caused by two effects: ({\it i}) star (D) is
near the onset of dynamical instability and 
hence a small deviation from true equilibrium can induce large 
perturbations; ({\it ii}) the
conformal flatness approximation results in larger deviations from
true equilibrium for more relativistic configurations, which causes a
larger initial perturbation.  Apparently, the deviation caused by the
numerical truncation error and/or the conformal flatness approximation
stabilizes the configuration, and for $\Delta K=0$ the star oscillates
with an average value of the central density 
($\rho(r=0) \simeq 0.004$) slightly smaller than
the initial value $\rho_c \simeq 0.0047$.  This suggests that star (D)
with $\Delta K=0$ is a perturbed state of a true equilibrium star of
$\rho_c \simeq 0.004 \sim \rho_{\rm crit}$.  The results for star
(E) are similar to, but more pronounced than those for star (D),
suggesting that the initial configuration (E) may also be a
perturbation of a stable star with $\rho_c \simeq 0.004 \sim 
\rho_{\rm crit}$.

To determine the onset of dynamical instability more sharply, we
perform further simulations adopting $\Delta K/K=0.7\%$, 0.8\%, and
$0.9\%$ for star (D).  In Fig. 3, we show $\rho$ and $\alpha$ at $r=0$ 
as a function of $t/{\rm P}$ for star (D) for the various initial
depletion factors.  We find that for $\Delta K/K \leq 0.8\%$, the
stars behave similarly to $\Delta K=0$; i.e., the stars simply
oscillate with the average density $\simeq \rho_{\rm crit}$.  For $\Delta
K/K \geq 0.9\%$, however, the stars quickly collapse to a black hole.
We do not find any examples in which the stars oscillate with average 
densities larger than $0.0045 \agt \rho_{\rm crit}$.  This indicates
that equilibrium stars with $\rho_c \agt \rho_{\rm crit}$ are
dynamically unstable.  Although we cannot specify the onset of
dynamical instability with arbitrary precision, our present results
indicate that it nearly coincides with the onset of secular
instability.


\subsection{Final outcome of unstable collapse}

To study the final outcome of the gravitational collapse of rapidly
rotating neutron stars, we perform a number of numerical experiments
for several different initial conditions.

First, we evolve star (D) with $\Delta K/K = 1\%$ for different
spatial gauge conditions.  In Fig. 4, we show $\phi$ and $\alpha$ at
$r=0$ as a function of time for the AMD gauge (solid line), the
modified gauge of type I (dotted line) and type II (dashed line). As
stated in Sec. II.B, $\phi(r=0)$ increases quickly during the
gravitational collapse for the AMD gauge.  In this case, $\alpha(r=0)$
stops decreasing in the late phase of the collapse where $\phi(r=0)
\agt 1$, which is a numerical artifact.  This is probably caused by
the insufficient resolution around the black hole forming region.  For
the modified gauge conditions, $\alpha(r=0)$ smoothly approaches
zero. We note that $\alpha(r=0)$ ought to be independent of the
spatial gauge condition, so that the deviation of the AMD results from
the modified gauge condition results are a numerical artifact.  This
shows that the results for $t/{\rm P} \agt 1.4$ computed in the AMD
gauge condition is unreliable and indicates that the modification of
the AMD gauge condition is an appropriate strategy to overcome the
deterioration of the resolution in the late phase of the collapse.

In Fig. 5, we show the time variation of the total angular momentum of
the system.  Since the evolving system is nearly axisymmetric, the
angular momentum should be nearly conserved.  In all the three cases,
however, the angular momentum slowly decreases in the early phase,
which is caused by numerical dissipation at the stellar surface.  As
the collapsing star approaches a black hole, the angular momentum
changes quickly because the resolution becomes increasingly worse. In
the AMD gauge case, the error amounts to $\agt 5\%$, while in the
modified gauge cases, it is $\sim 1.5\%$ at the time when apparent
horizon is found at $t \sim 1.4{\rm P}$ (see Fig. 6).  This is further
evidence that the modified gauge conditions are better suited for
simulations of black hole formation.

It should be noted that even with the modified gauge conditions, the
resolution becomes too poor to perform accurate simulations for times
exceeding $t/{\rm P} \agt 1.5$. This is because the metric $\tilde
\gamma_{ij}$ becomes very spiky around the apparent horizon (i.e.,
because of horizon throat stretching). To perform simulations for
times much later than horizon formation special computational tools
are necessary, probably including apparent horizon boundary
conditions~\cite{Suen}.

In Fig. 6, we show snapshots of density contour lines for the density
$\rho_*$ and the velocity field for $v^i(=u^i/u^0)$ in the equatorial
and $y=0$ planes. The results are obtained in the modified gauge
condition of type I. It is found that after about 1.4 orbital periods
almost all the matter has collapsed to a black hole.  In Fig. 7, we
show the fraction of the rest mass inside a coordinate radius $r$,
defined as
\beq
{M_*(r) \over M_*}=
{1 \over M_*} \int_{|x^i| < r} d^3x \rho_*. 
\eeq
$R_e$ denotes the coordinate axial length in the equatorial plane at
$t=0$ (see Table I).  Note that at $t \sim 1.4{\rm P}$, the apparent
horizon is located at $r\simeq 0.2R_e$.  Thus, almost all the matter
(more than $99\%$) has been absorbed by the black hole by that time.
Although Fig. 6 shows that a small fraction of the matter has not yet
been swallowed by the black hole, the matter which stays inside $r
\alt R_e \sim 5M_g$ will ultimately have to fall in.  This is, because
the radius of the innermost stable circular orbit (ISCO) is $R^{\rm
SS}_{\rm ISCO} \sim 5 M_g$ for a (nonrotating) Schwarzschild black hole
in our gauge.  The collapse of rotating neutron stars with $J/M_g^2
\sim 0.6 <1$ leads to moderately rotating Kerr black holes, for which
$R^{\rm SS}_{\rm ISCO}$ is an adequate approximation to the ISCO.
This fact already suggests that no disk will form around the black
hole.

The same reason also suggests why no massive disk forms during the
collapse: the equatorial radius $R_e$ is initially less than $5M_g$,
and hence inside the radius which will become the ISCO of the final
black hole.

Next, we evolve the initial configuration (A) depleting the pressure
by various amounts, which may provide a model for sudden phase
transitions inside neutron stars~\cite{Bethe,Glendenning}.  In Fig.~8,
we show $\rho$ and $\alpha$ at $r=0$ as a function of time for $\Delta
K/K=0, 1\%$, $5\%$, and $10\%$~\cite{footnote}.  When the depletion
factor is less than $5\%$, the star simply oscillates, but for $\Delta
K/K=10\%$ the star collapses dynamically.  Note that depleting the
pressure by $10\%$ is approximately equivalent to increasing the
gravitational mass by $5\%$ according to the scaling relation for the
polytropic stars of $\Gamma=2$ (see Eq.~(\ref{rescale})).  Since the
gravitational mass for star (A) is about $3\%$ less than the maximum
allowed mass, it is quite reasonable that this star collapses.  In the
following two simulations, we focus on evolutions of star (A) with
$\Delta K/K=10\%$.

In order to test if nonaxisymmetric (bar-mode) perturbations have
enough time to grow appreciably during the gravitational collapse, we
excite such a perturbation by modifying the initial density profile
$\rho_*$ according to~\cite{footnote}
\beq
\rho_*=(\rho_*)_{0}\Bigl(1+0.3{x^2-y^2 \over R_e^2}\Bigr),
\eeq
where $(\rho_*)_{0}$ denotes the density profile of 
star (A) in the unperturbed state. 

In Figs.~9 and 10, we show snapshots of density contour lines for
$\rho_*$ and the velocity field for $v^i$ in the equatorial and $y=0$
planes for the above axisymmetric and nonaxisymmetric initial
conditions.  For these simulations we adopted the modified gauge
condition of type II.  In Fig.~11, we also show $M_*(r)/M_*$ as a
function of time for these cases.  We again find that irrespective of
the initial perturbation, almost all the matter collapses into the
black hole without any massive disk or ejecta around the black hole.
Again, this is a consequence of the stars being sufficiently compact
that almost all the matter ends up inside the ISCO of the final black
hole.  Note that the star with the nonaxisymmetric perturbation
evolves very similarly to the unperturbed, axisymmetric star, showing
that the dynamical collapse does not leave the nonaxisymmetric
perturbation enough time to grow appreciably during the collapse.
Again, this can be understood quite easily from the following
heuristic (and Newtonian) argument.  Star (A) has an initial
equatorial radius of $R_e \sim 5.5 M_g$, and can therefore shrink by
less than a factor of 3 before a black hole forms.  Its initial value
of $T/|W|$ is about 0.09 (see Table I).  Since $T/|W|$ scales
approximately with $R^{-1}$, it can just barely reach the critical
value $(T/|W|)_{\rm dyn} \sim 0.27$ for dynamical instability before a
black hole forms.  It is therefore not surprising that we do not find
dynamically growing axisymmetric perturbations.  Note that the star
does reach the critical value for secular instability to bar
formation (which may be as small as $(T/|W|)_{\rm sec} \sim 0.1$
for very compact configurations, see~\cite{SF}), so that viscosity or
emission of gravitational waves could drive the star unstable.
However, this mode would grow on the corresponding dissipative
timescale, which is much longer than the dynamical timescale of the
collapse.

In order to make these statements about nonaxisymmetric growth more
quantitative, we compare the quantities
\beq
2{x_{\rm rms}-y_{\rm rms} \over x_{\rm rms}+y_{\rm rms}}
\eeq
for the perturbed and unperturbed evolutions in Fig.~12.  Here, 
$x^i_{\rm rms}$ denotes the mean square axial length defined as 
\beq
x^i_{\rm rms}=\biggl[
{1 \over M_*}\int d^3x \rho_* (x^i)^2\biggr]^{1/2}.
\eeq
The figure shows very clearly that the axial ratio oscillates for the
perturbed evolution, but does not grow on the dynamical timescale 
of the collapse.

Finally, we model a scenario in which a small amount of matter
accretes onto a stable star resulting in destabilization of the
star. As the stable star, we again adopt configuration (A) and to
model the matter accretion we modify the initial density distribution
according to~\cite{footnote}
\beq
\rho_*=(\rho_*)_{0}\Bigl(1+0.5{r^2 \over R_e^2}\Bigr),
\eeq
with all the matter moving with the same initial angular velocity.
Most of the enhancement is in the outer region, which mimics the
effect of accretion.  In this case, the total rest mass is about
$9.5\%$ larger than that of star (A), so that the mass is larger than
the maximum allowed mass along the sequence of rotating neutron stars.
The value of $J/M_g^2$ is nearly unchanged.
Note that we do not reduce the pressure for these simulations.  We
again evolve the star using the modified gauge condition of type II.

The star again evolves very similarly to those in the previous two
cases. As an example, we show in Fig~13 $M_*(r)/M_*$ as a function of
time, which is similar to the results in Fig.~11.  The apparent
horizon forms at $t \simeq 0.89{\rm P}$ and $r \simeq 0.2R_e$.  We
again find that almost all the matter collapses into the black hole
without forming a massive disk around the black hole.

\section{Summary and conclusion}

We perform fully relativistic, 3D hydrodynamic simulations of
supramassive neutron stars rigidly rotating at the mass-shedding
limit.  We study the dynamical stability of such stars close to the
onset of secular instability and follow the collapse to rotating black
holes.

Our results suggest that the onset of dynamical, radial instability is
indeed close to the onset of secular instability, as expected from the
coincidence of the secular and dynamical instability in nonrotating,
spherical stars.

In all our simulations, nearly all the matter is consumed by the
nascent black hole by the time the calculation stops, and we do not
find any evidence for a formation of a massive disk or any ejecta.
Since we are considering maximally rotating neutron stars at the
mass-shedding limit, and since the formation of a disk is even less
likely for more slowly rotating stars, we conclude that such disks
quite generally do not form during the collapse of unstable, uniformly
rotating neutron stars.  This also includes stars which are
destabilized by pressure depletion (as, for example, by a nuclear
phase transition), or by mass accretion.

We also find that during the collapse to a black hole, nonaxisymmetric
perturbations do not have enough time to grow appreciably.  

Both these findings can be understood quite easily from heuristic
arguments.  The initial equilibrium configurations are sufficiently
compact, typically $R_e \lesssim 6 M_g$, so that most of the matter
already starts out inside the radius which will become the ISCO of the
final black hole.  Therefore it is very unlikely that a stable,
massive disk would form.  Also, the star can only contract by about a
factor of three before a black hole forms. Hence $T/|W|$, which
approximately scales with $R^{-1}$, can only increase by
about a factor of three over its initial value of $(T/|W|)_{\rm init}
\sim 0.09$, and only barely reaches the critical value of
dynamical instability for bar formation $(T/|W|)_{\rm dyn} \sim 0.27$.
It is therefore not surprising that we do not see a dynamical growth
of nonaxisymmetric perturbations.  We expect that these results hold
for any moderately stiff equation of state, for which the
corresponding critical configurations are similarly compact.

The study reported here focuses on {\em uniformly} rotating neutron stars,
for which we adopt a moderately {\em stiff} equation of state and consider a
configuration which is moderately compact initially ($R/M_g \sim 6$).  
We speculate that for two alternative scenarios the results may be quite 
different, even qualitatively, both as far as the formation of a disk 
and the growth of nonaxisymmetric perturbations are concerned.  

For {\em differentially} rotating neutron stars, which are the likely outcome
of the merger of binary neutron stars \cite{BSS}, $T/|W|$ may take larger 
values than for rigidly rotating neutron stars.  It is therefore
possible that such stars might develop dynamical bar mode instabilities.

Rotating supermassive stars (with masses $M \gtrsim 10^5 M_{\odot}$)
or massive stars on the verge of supernova collapse are subject to the
same dynamical instabilities, but are characterized by very {\em soft}
equations of state ($\Gamma \sim 4/3$) and initial configurations
which are nearly Newtonian (see~\cite{ST,BS}).  Such stars therefore
reach the critical value $(T/|W|)_{\rm dyn}$ for bar mode formation
far outside the horizon radius.  Moreover, $R/M$ is very large
initially, so that a disk may easily form (compare the discussion
in~\cite{BS}).

We will treat the collapse of both differentially rotating neutron 
stars and supermassive stars in future papers.

\acknowledgments

Numerical computations were performed on the FACOM VPP 300R and VX/4R
machines in the data processing center of the National Astronomical
Observatory of Japan.  This work was supported by NSF Grants AST
96-18524 and PHY 99-02833 and NASA Grant NAG5-7152 at the University
of Illinois.  M.S. gratefully acknowledges support by Grant-in-Aid
(Nos.~08NP0801 and 09740336) of the Japanese Ministry of Education,
Science, Sports and Culture, and JSPS (Fellowships for Research Abroad).

\vskip 5mm
\noindent 
{\bf Table I.~} The list of the central density $\rho_c$, 
total rest mass $M_*$, gravitational mass $M_g$, 
angular momentum in units of $M_g^2$ ($J/M_g^2$), $T/W$, 
rotation period ${\rm P}$, and 
coordinate length of the semi-major axis $R_e$  for 
rotating neutron stars at mass-shedding limits in the 
conformal flat approximation. The gauge invariant definition of 
$T/W$ is the same as that in Ref.~\cite{CST2}. 
The units of mass, length and time are 
$M_{\odot}$, $1.477$km, and $4.927\mu$sec, respectively.

\vskip 5mm
\noindent
\begin{center}
\begin{tabular}{|c|c|c|c|c|c|c|c|c|} \hline
\hspace{1mm} $\rho_{c}(10^{-3})$ \hspace{1mm} &
\hspace{1mm} $M_*$ \hspace{1mm} & \hspace{1mm} $M_g$\hspace{1mm} 
& $J/M_g^2$  & $T/W$ 
& \hspace{1mm} ${\rm P} $ \hspace{1mm} 
& \hspace{1mm} $R_e$ \hspace{1mm} & Model \\ \hline
$2.77 $& 1.580 & 1.452 & 0.598&0.087 &163& 8.064& (A) \\ \hline
$3.38 $& 1.628 & 1.484 & 0.586&0.085 &148& 7.820& (B) \\ \hline
$3.98 $& 1.646 & 1.496 & 0.574&0.083 &137& 7.365& (C) \\ \hline
$4.68 $& 1.645 & 1.494 & 0.563&0.080 &127& 6.934& (D) \\ \hline
$5.43 $& 1.630 & 1.481 & 0.553&0.078 &120& 6.566& (E) \\ \hline
\end{tabular}
\end{center}

\begin{figure}[t]
\epsfxsize=3in
\leavevmode
\epsffile{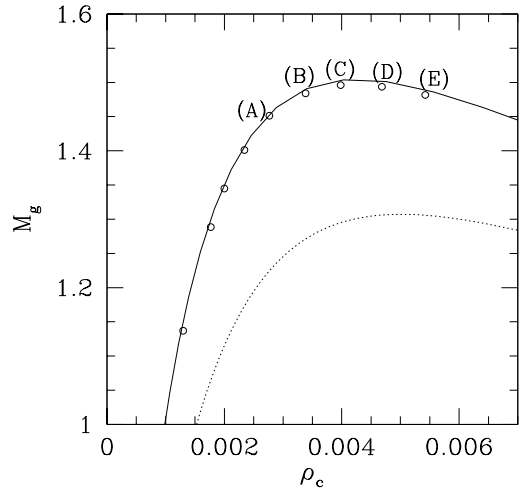}
\caption{
The gravitational mass $M_g$ as a function of the central density
$\rho_c$ for rotating stars with $\Gamma=2$ and $K=200/\pi$.  The
solid and dashed lines denote exact solutions for sequences of
rotating stars at the mass-shedding limit and spherical stars.  The
open circles denote the sequence of rotating stars at the
mass-shedding limit as obtained from the conformal flatness
approximation.  The configurations that we adopt as initial data for
dynamical evolution calculations in this paper are marked with
(A)--(E).  }
\end{figure}

\clearpage

\begin{figure}[t]
\epsfxsize=3in
\leavevmode
\epsffile{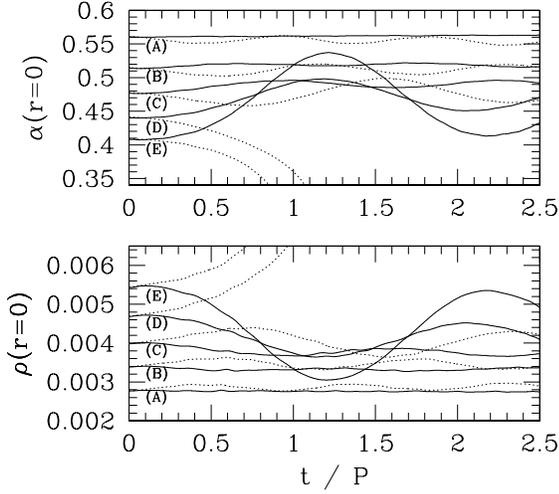}
\caption{$\alpha$ and $\rho$ at $r=0$ as a function of $t/{\rm P}$ 
in the evolution of stars (A)--(E). 
The solid and dotted lines denote the results for $\Delta K=0$ 
and $\Delta K/K=1\%$, respectively. 
}
\end{figure}

\begin{figure}[t]
\epsfxsize=3in
\leavevmode
\epsffile{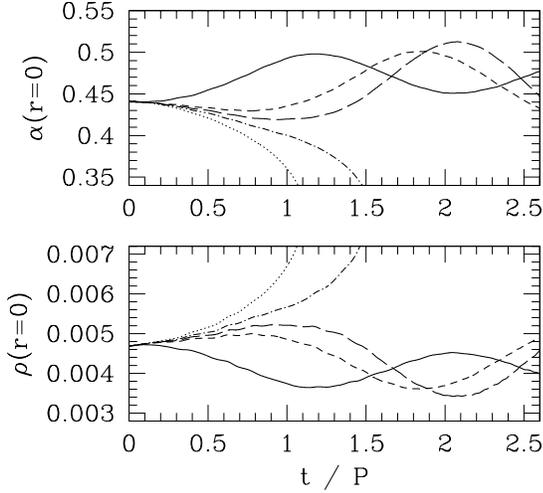}
\caption{$\alpha$ and $\rho$ at $r=0$ as a function of $t/{\rm P}$ 
during the evolution of star (D) for various $\Delta K/K$. 
The solid, dashed, long dashed, dotted-dashed, and dotted 
lines denote the results for $\Delta K/K=0$, 0.7\%, 0.8\%, 0.9\% and 
$1\%$. 
}
\end{figure}

\begin{figure}[t]
\epsfxsize=3in
\leavevmode
\epsffile{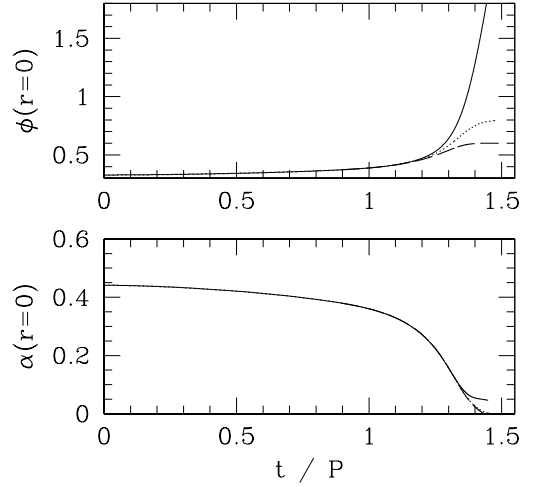}
\caption{$\phi$ and $\alpha$ at $r=0$ as a function of $t/{\rm P}$ 
during the collapse of star (D) with $\Delta K/K=1\%$ 
for the AMD gauge (the solid line), the modified gauge of type I 
(the dotted line) and of type II (the dashed line), 
respectively. 
}
\end{figure}

\begin{figure}[t]
\epsfxsize=3in
\leavevmode
\epsffile{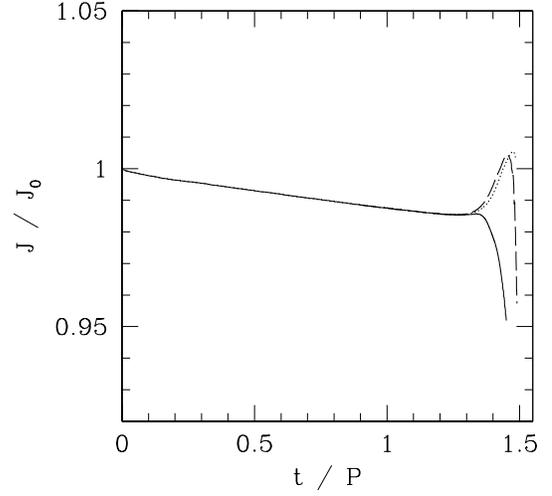}
\caption{Same as Fig.~4, but for the angular momentum 
$J/J_0$ as a function of $t/{\rm P}$. 
Here, $J_0$ is the angular momentum of the system at $t=0$.}
\end{figure}

\clearpage

\begin{figure}[t]
\begin{center}
\epsfxsize=2.5in
\leavevmode
\epsffile{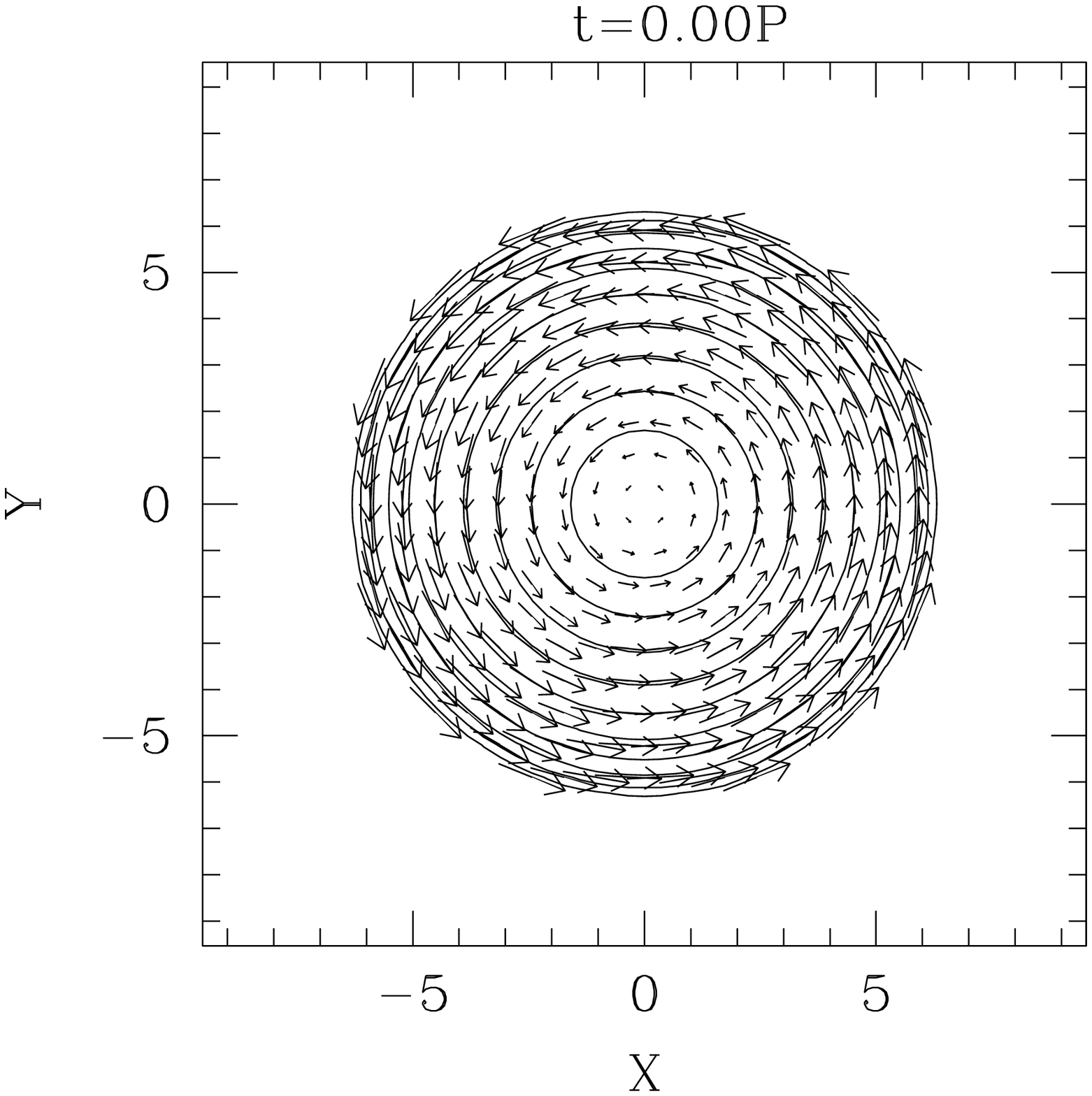}
\epsfxsize=2.5in
\leavevmode
\epsffile{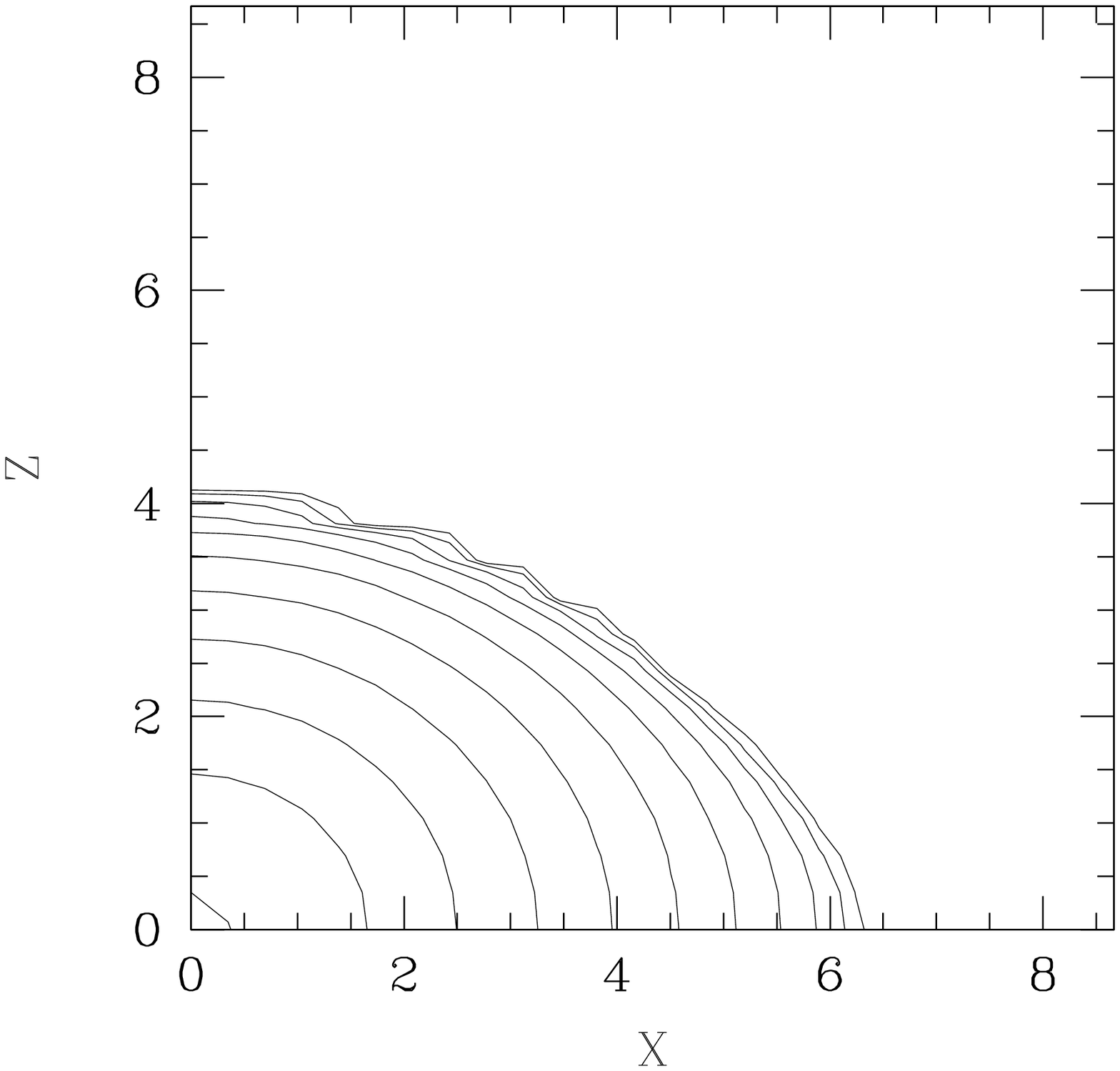}\\
\epsfxsize=2.5in
\leavevmode
\epsffile{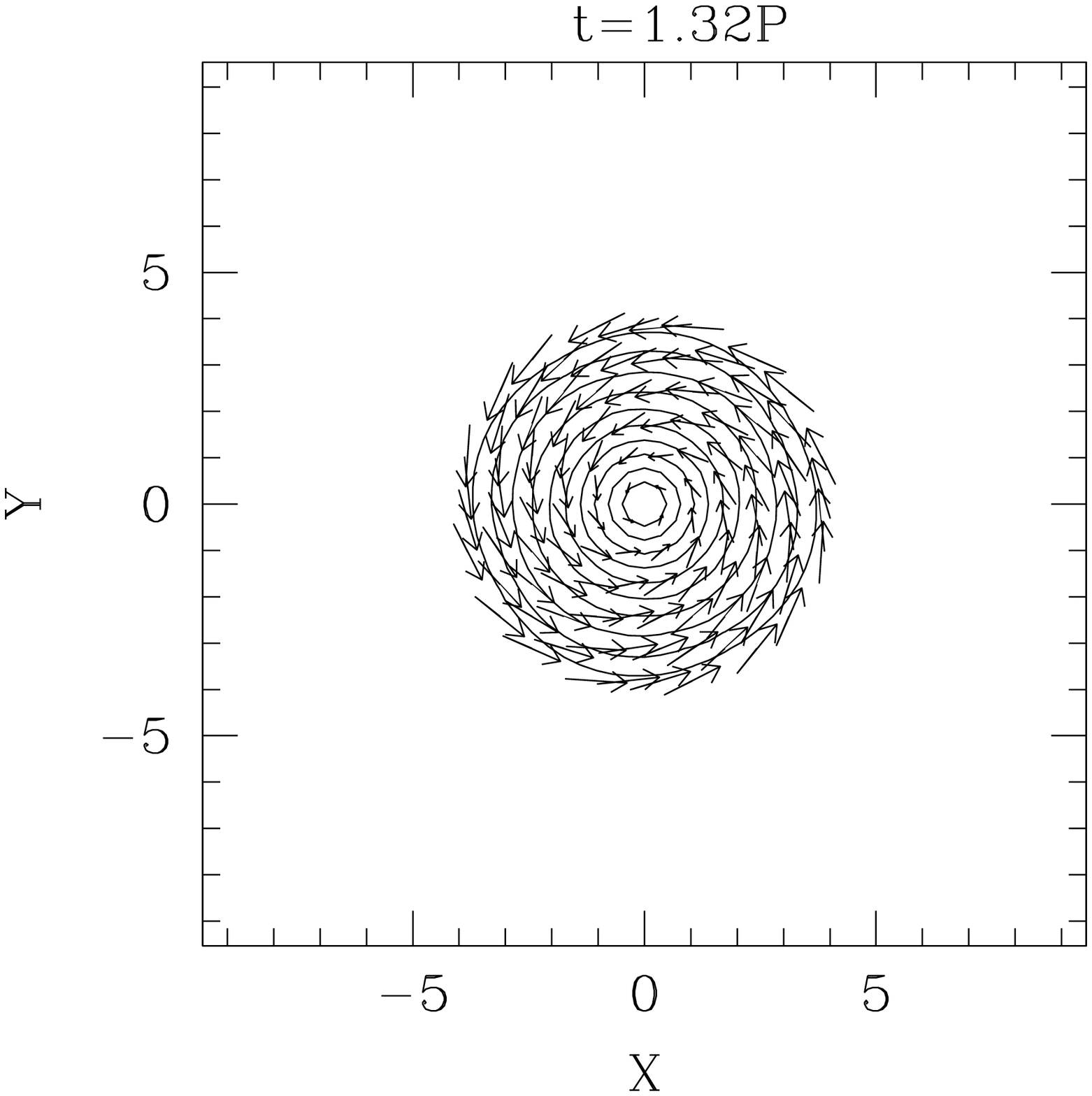}
\epsfxsize=2.5in
\leavevmode
\epsffile{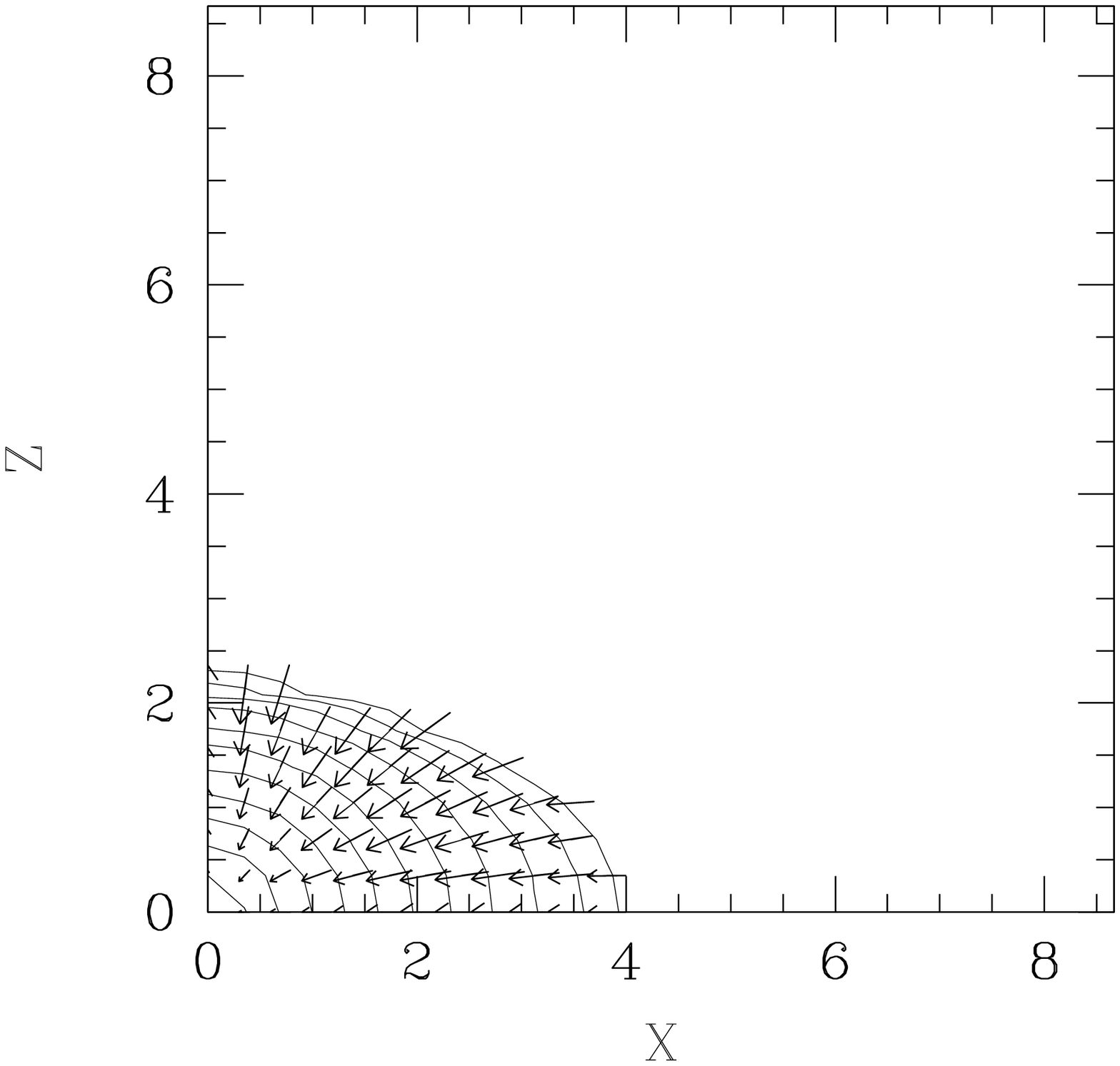}\\
\epsfxsize=2.5in
\leavevmode
\epsffile{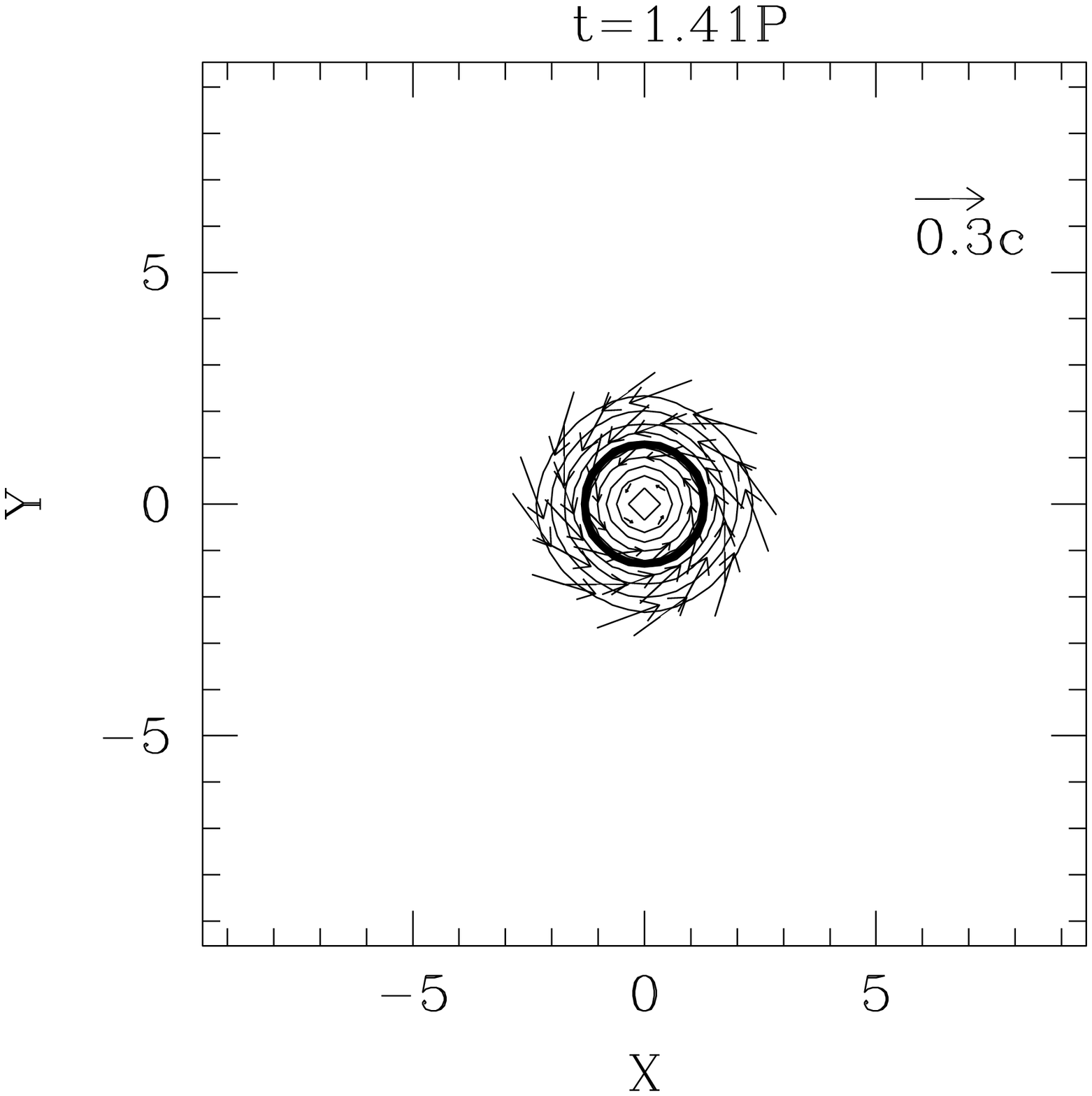}
\epsfxsize=2.5in
\leavevmode
\epsffile{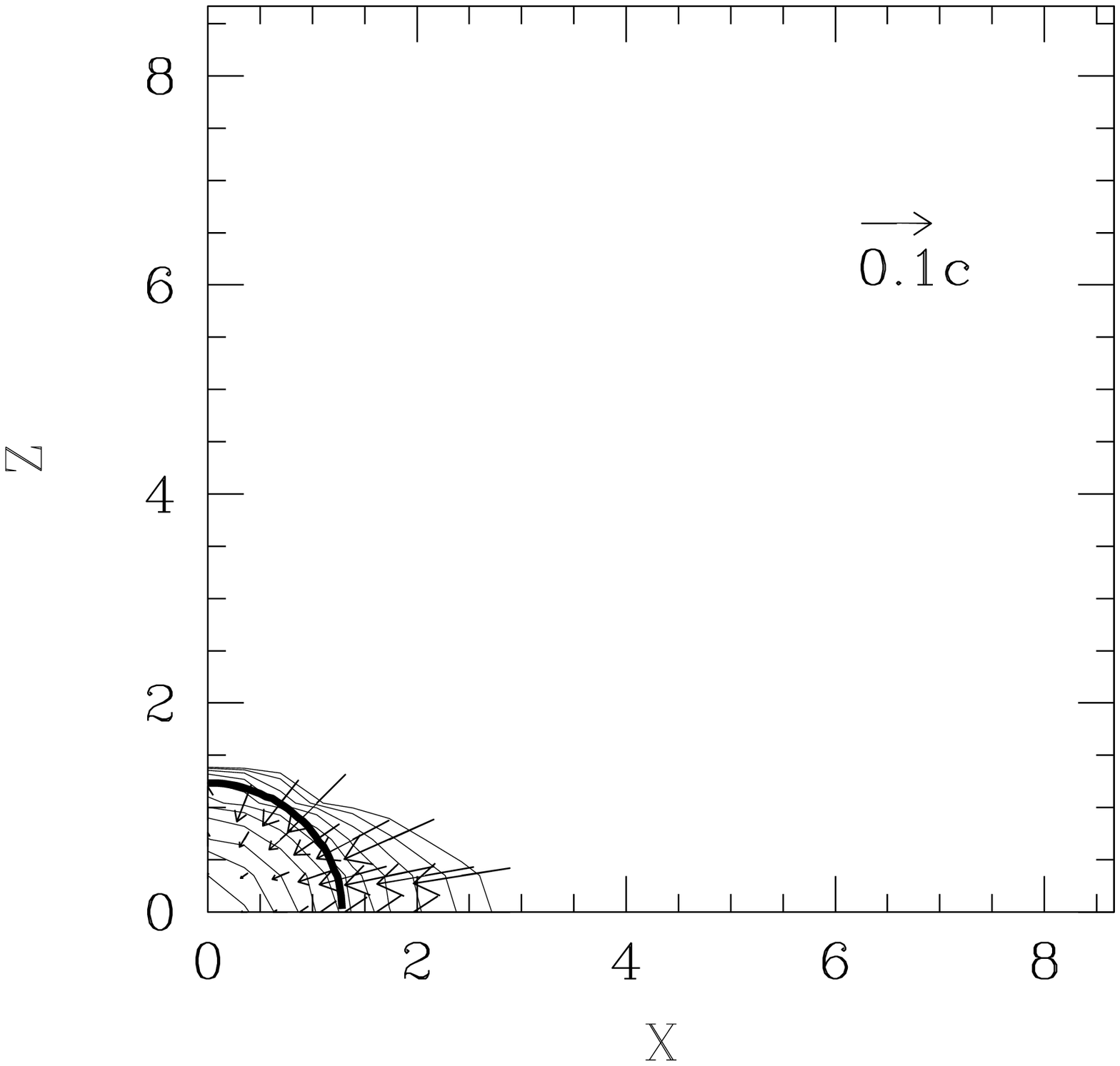}
\end{center}
\caption{Snapshots of density contours for $\rho_*$ and 
the velocity flow for $(v^x,v^y)$ in the equatorial plane (left) and
in the $y=0$ plane (right) for the collapse of star (D) with $\Delta
K/K=1\%$ (evolved with type I modified AMD gauge). The contour lines
are drawn for $\rho_*/\rho_{*~c}=10^{-0.3j}$ for $j=0,1,2,\cdots,10$
where $\rho_{*~c}$ is 0.034, 0.64 and 2.04 for the three different
times.  The lengths of arrows are are normalized to $0.3c$ (left) and
$0.1c$ (right).  The thick solid lines denote the location of the
apparent horizon. }
\end{figure}

\clearpage

\begin{figure}[t]
\epsfxsize=3in
\leavevmode
\epsffile{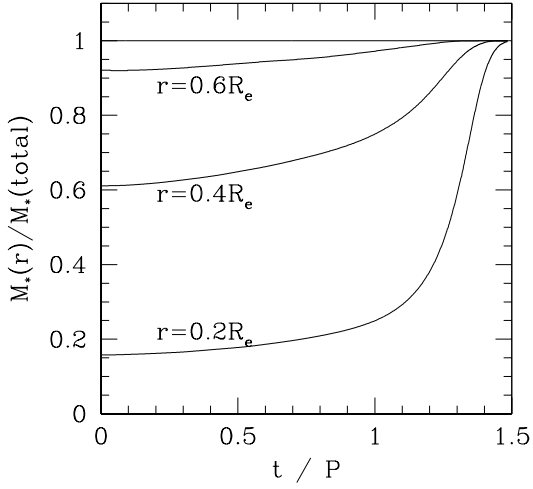}
\caption{Fraction of the 
rest mass inside a coordinate radius $r$ as a function of $t/{\rm P}$ 
for star (D) with $\Delta K/K=1\%$ (evolved with type I modified AMD gauge). 
$R_e$ denotes the initial coordinate length of the semi major axis. }
\end{figure}

\begin{figure}[t]
\epsfxsize=3in
\leavevmode
\epsffile{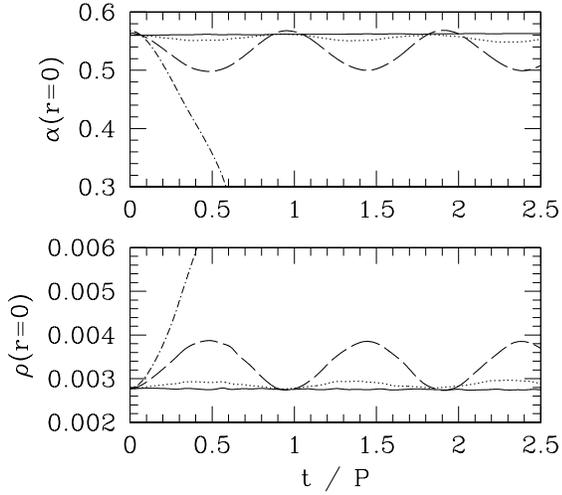}
\caption{$\alpha$ and $\rho$ at $r=0$ as a function of $t/{\rm P}$ 
for star (A) of various initial pressure depletion factors $\Delta
K/K$.  The solid, dotted, dashed and dotted-dashed lines denote the
cases where $\Delta K/K=0$, 1\%, 5\%, and $10\%$, respectively.}
\end{figure}

\clearpage
\begin{figure}[t]
\begin{center}
\epsfxsize=2.5in
\leavevmode
\epsffile{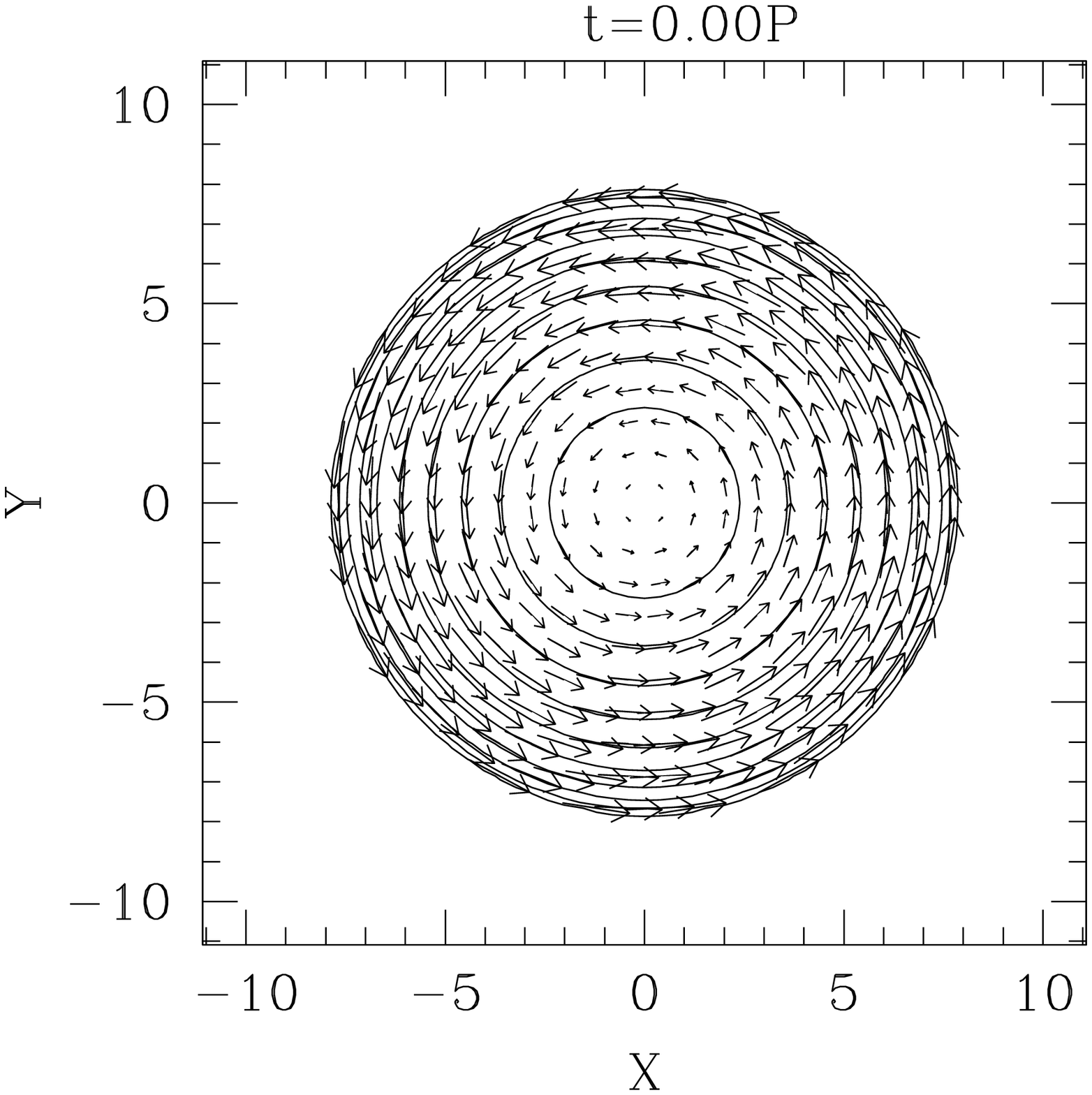}
\epsfxsize=2.5in
\leavevmode
\epsffile{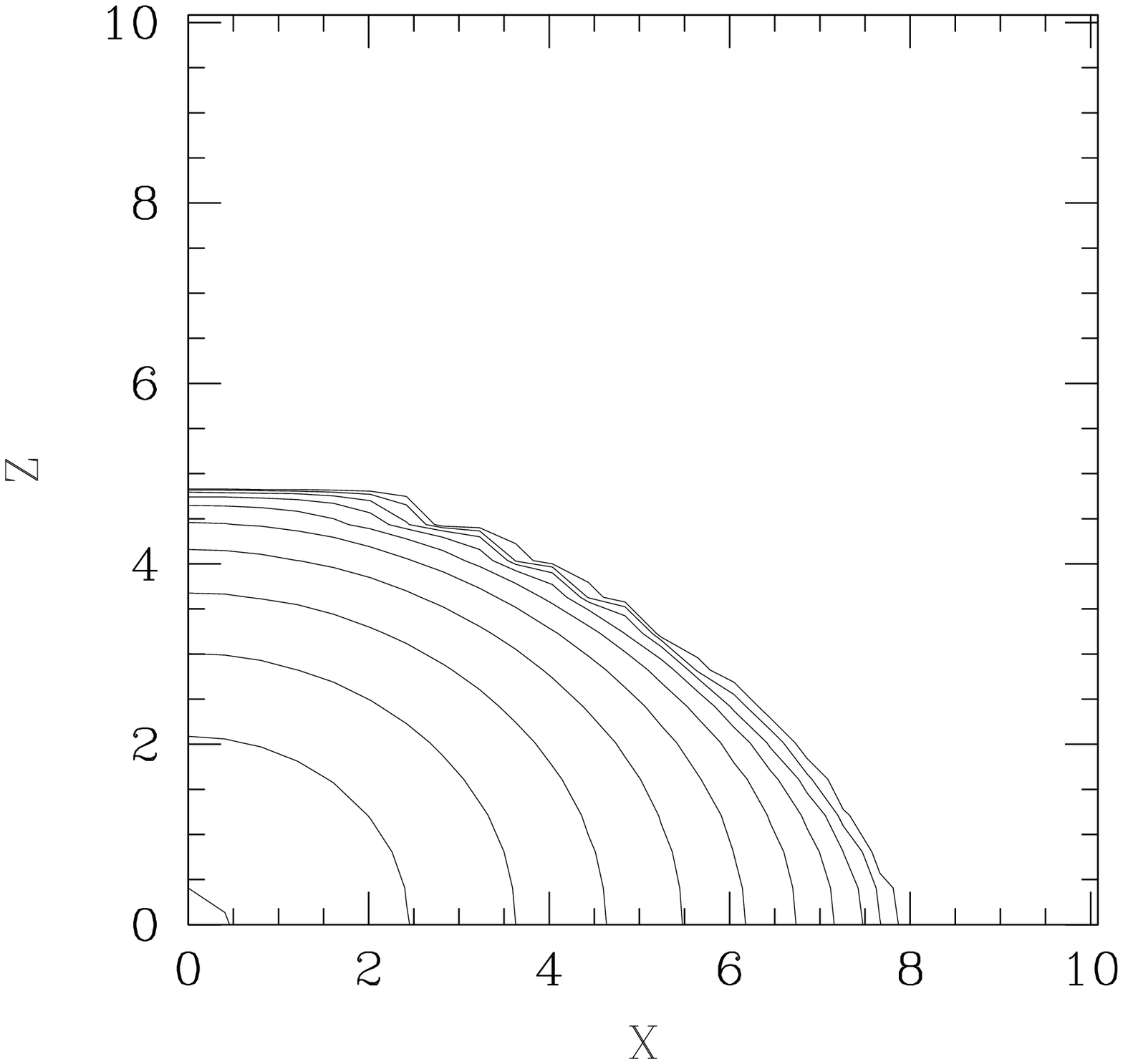}\\
\epsfxsize=2.5in
\leavevmode
\epsffile{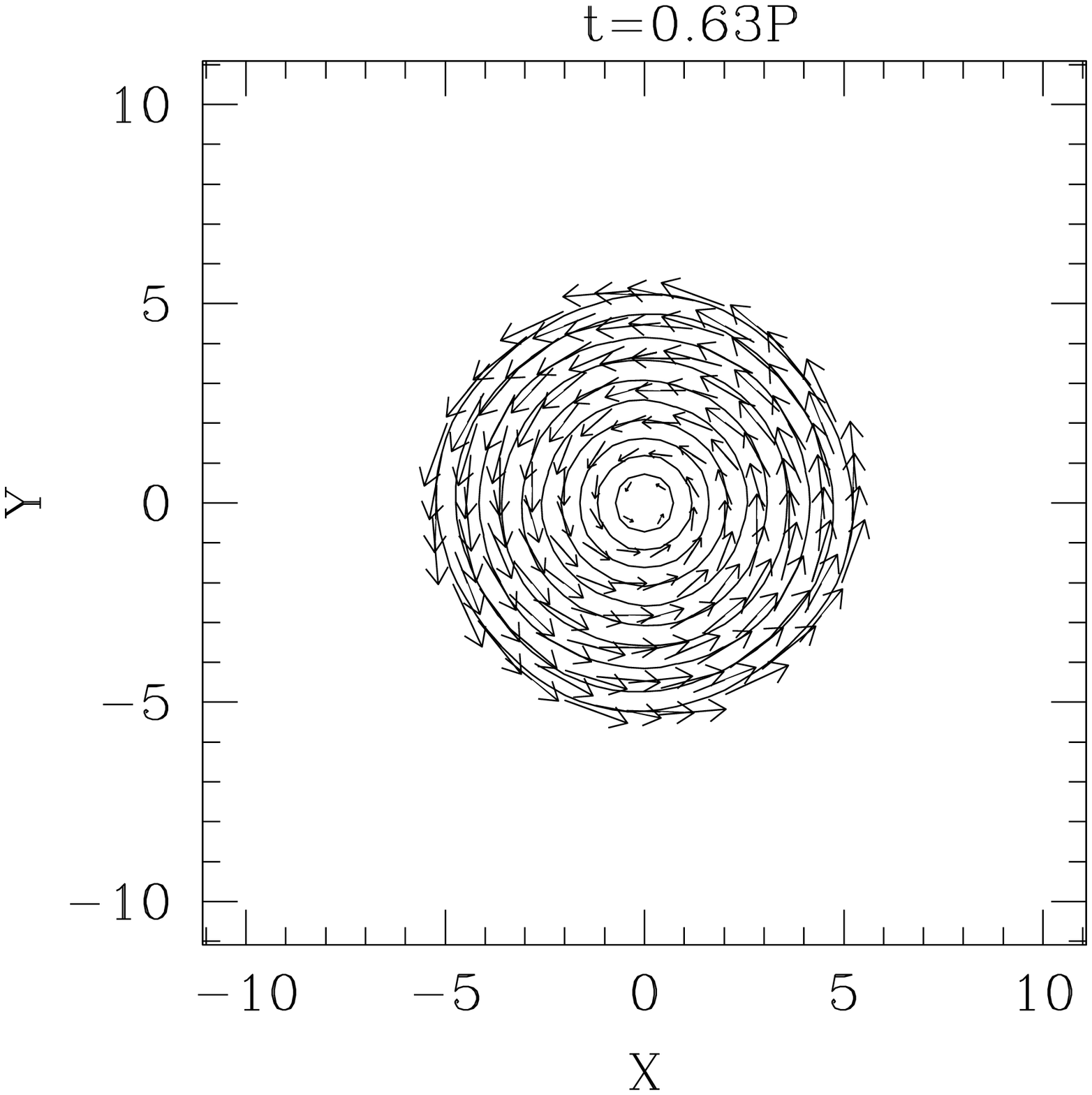}
\epsfxsize=2.5in
\leavevmode
\epsffile{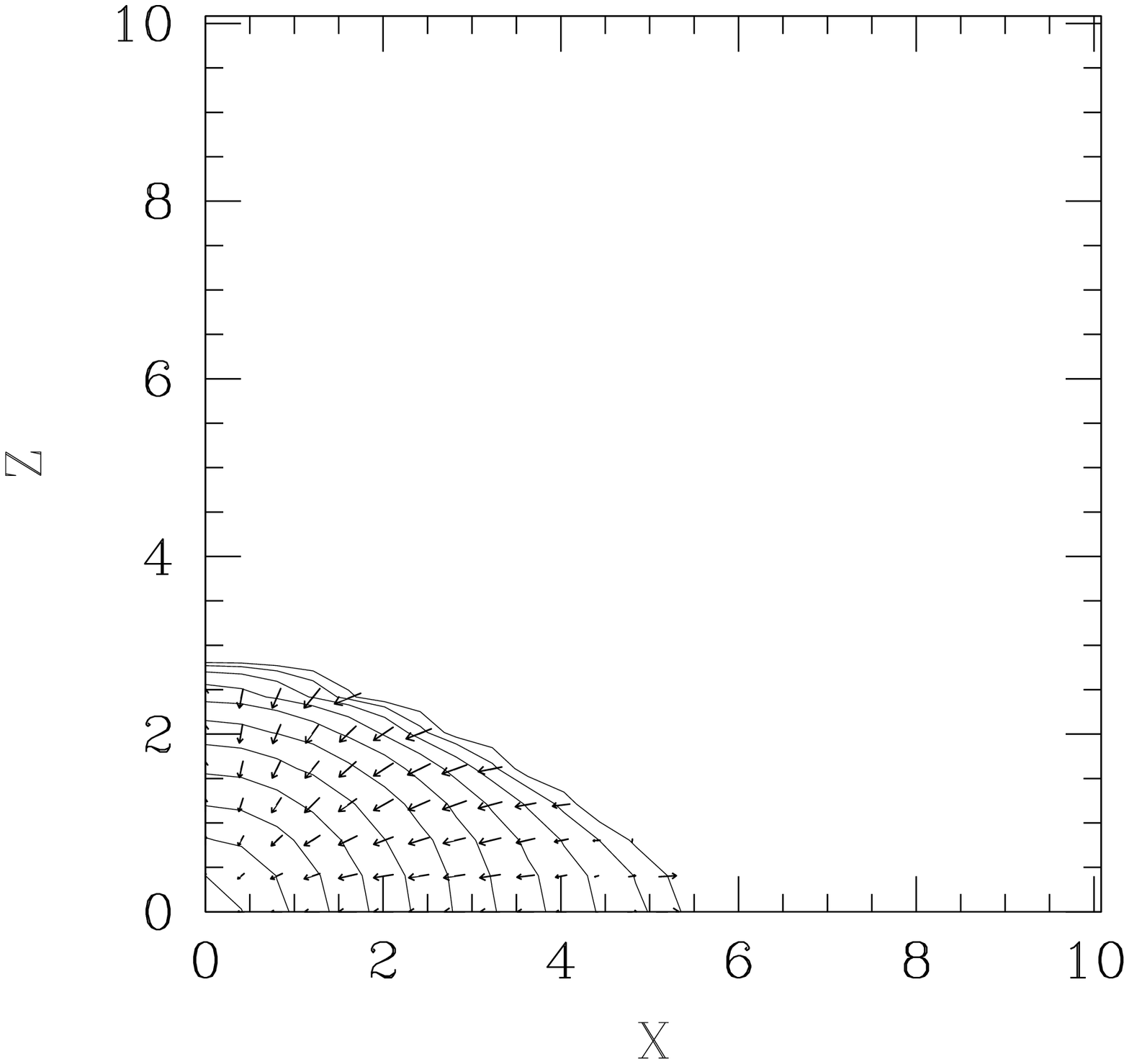}\\
\epsfxsize=2.5in
\leavevmode
\epsffile{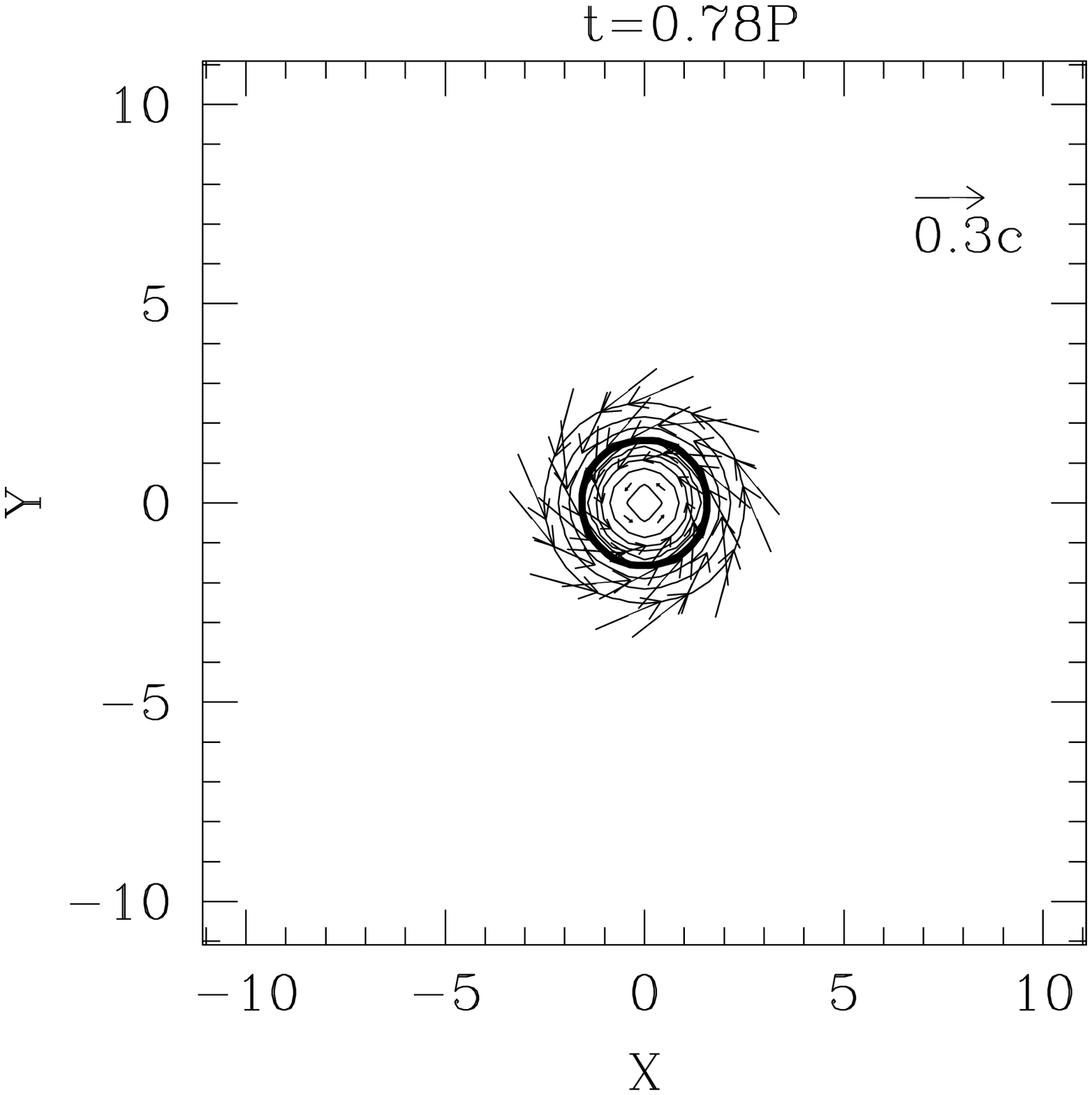}
\epsfxsize=2.5in
\leavevmode
\epsffile{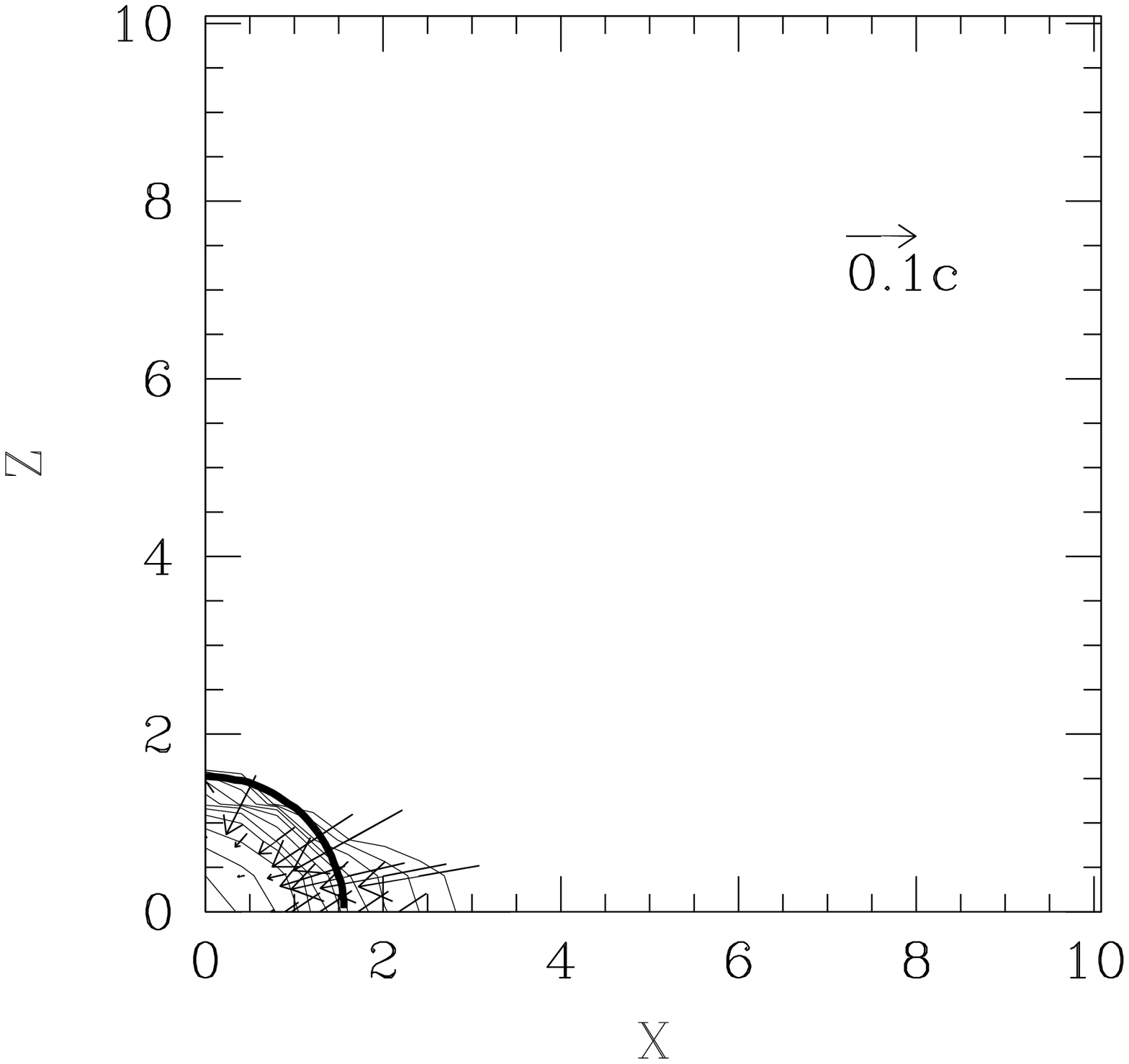}
\end{center}
\caption{Same as Fig. 6, but for star (A) with $\Delta K/K=10\%$. 
The contour lines are drawn for $\rho_*/\rho_{*~c}=10^{-0.3j}$ for
$j=0,1,2,\cdots,10$ where $\rho_{*~c}$ is 0.012, 0.20, and 1.01 for
the three different times.  }
\end{figure}

\clearpage
\begin{figure}[t]
\begin{center}
\epsfxsize=2.5in
\leavevmode
\epsffile{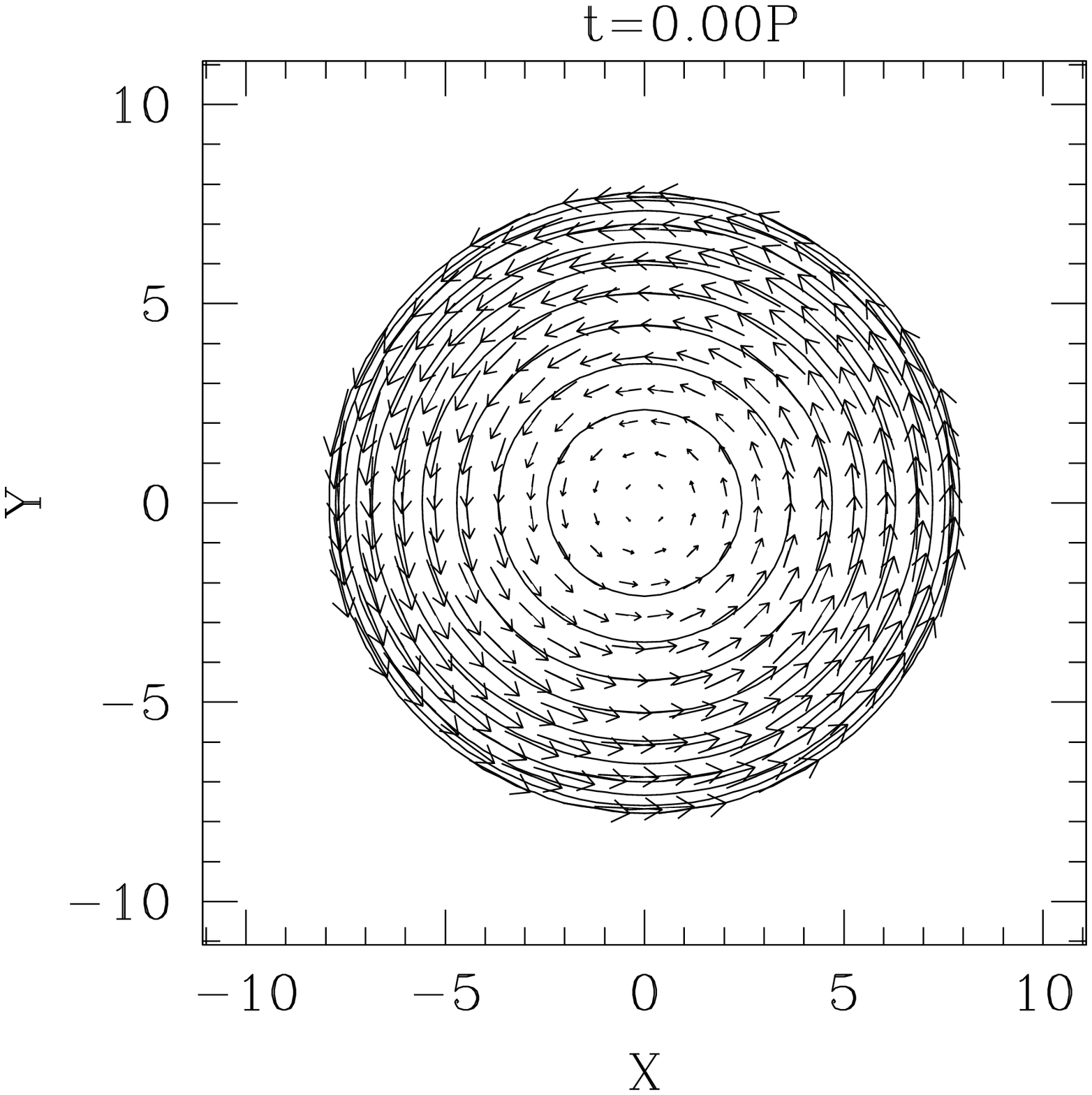}
\epsfxsize=2.5in
\leavevmode
\epsffile{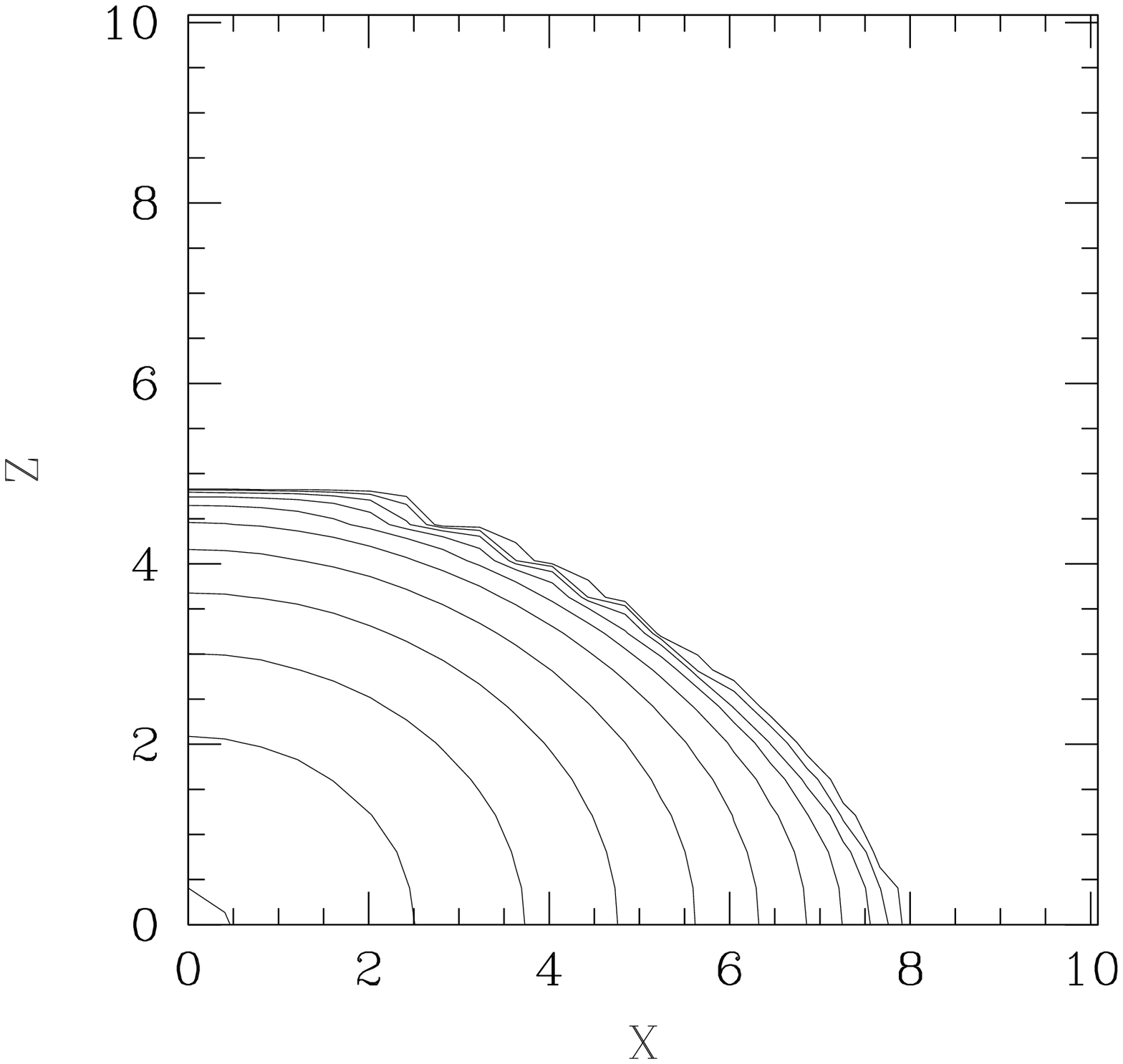}\\
\epsfxsize=2.5in
\leavevmode
\epsffile{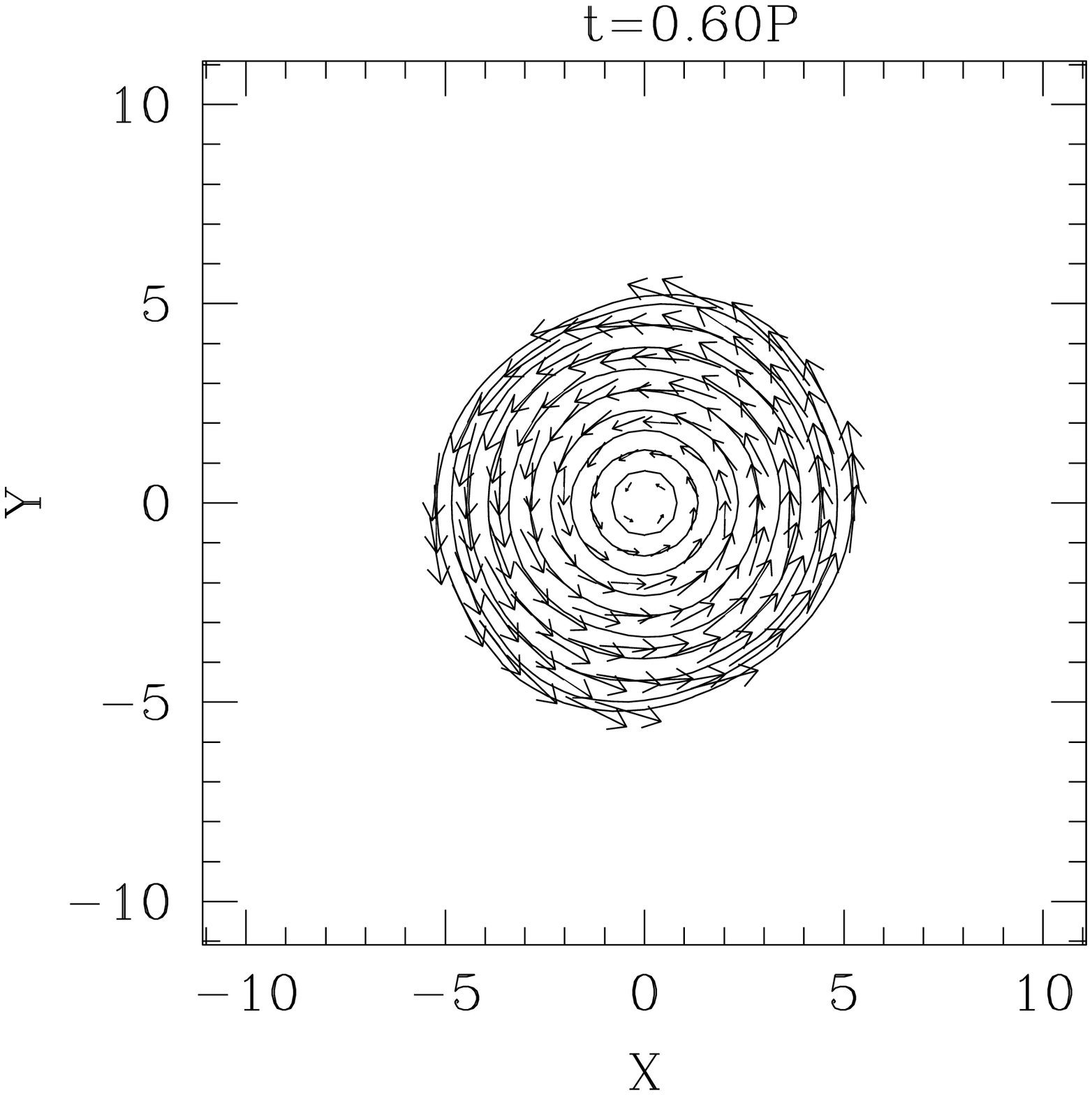}
\epsfxsize=2.5in
\leavevmode
\epsffile{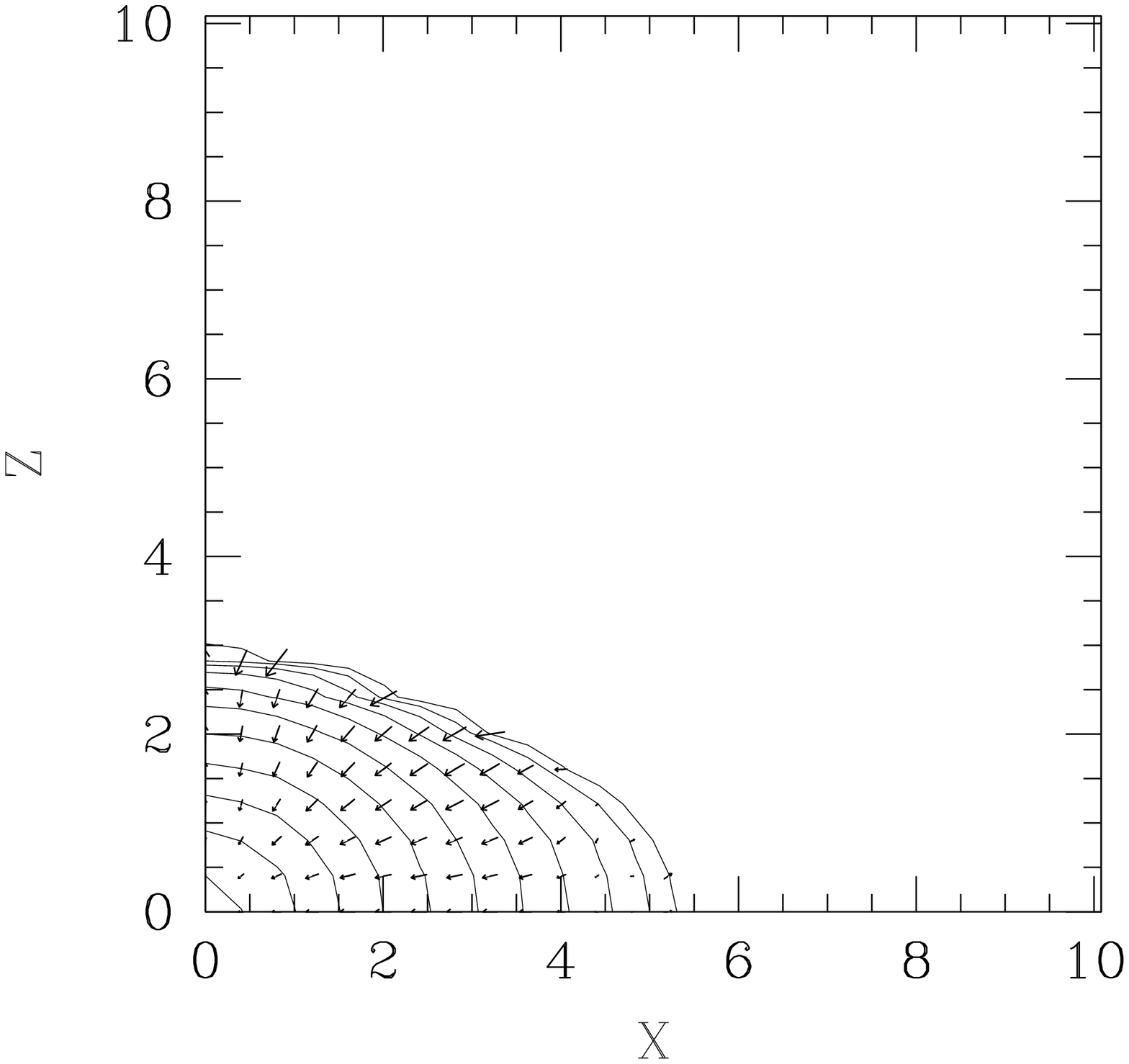}\\
\epsfxsize=2.5in
\leavevmode
\epsffile{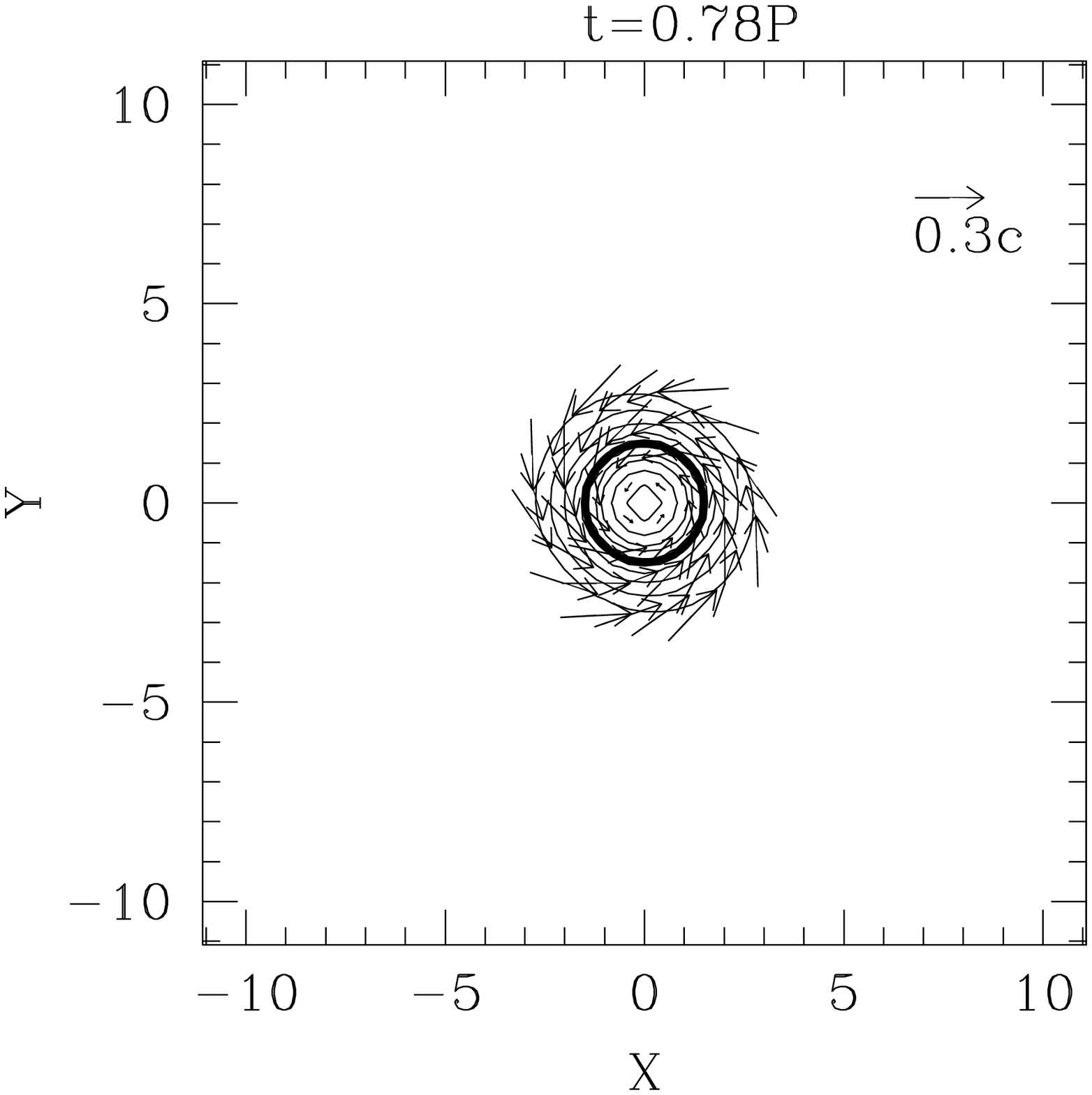}
\epsfxsize=2.5in
\leavevmode
\epsffile{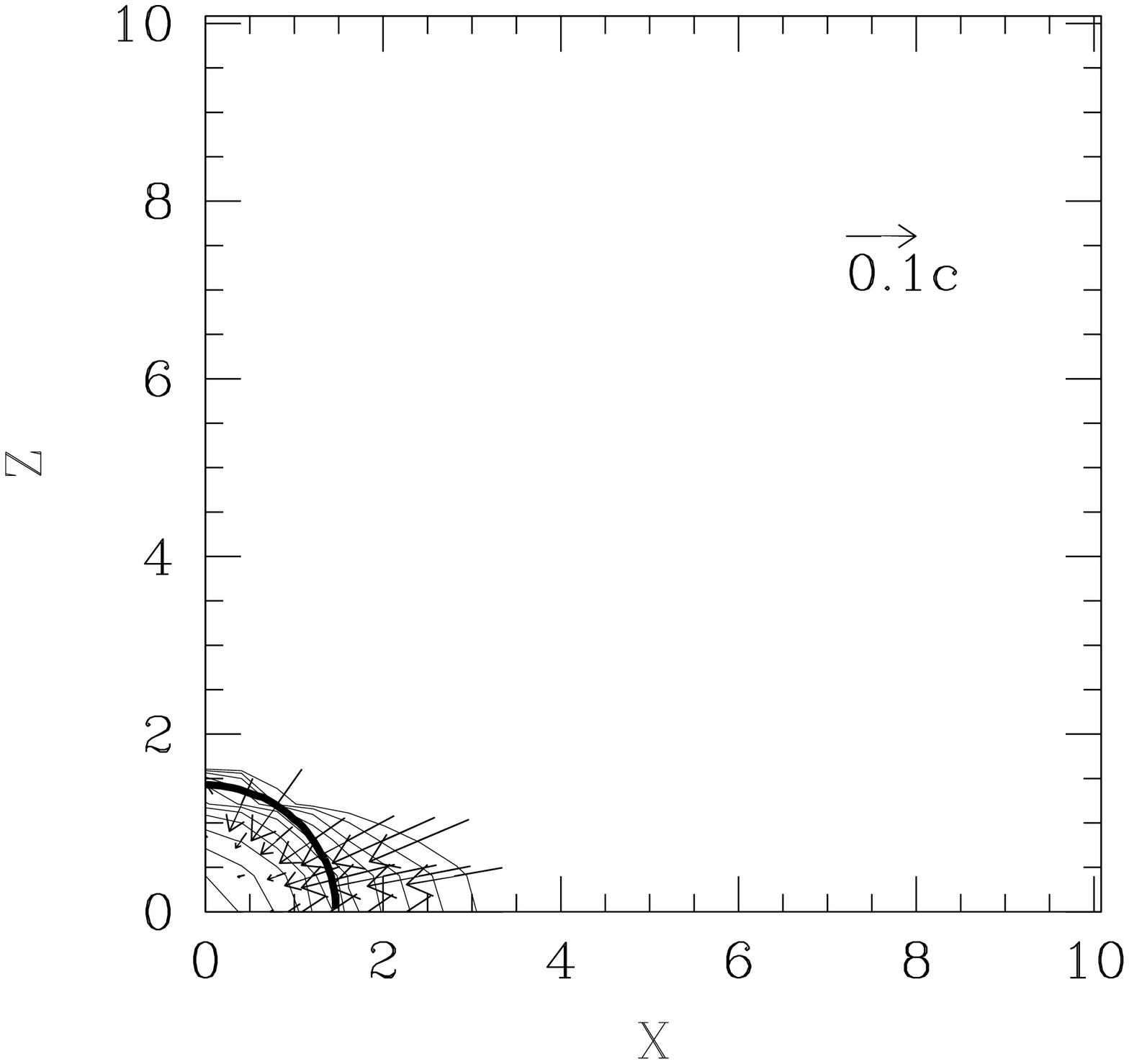}
\end{center}
\caption{Same as Fig. 6, 
but for star (A) with $\Delta K/K=10\%$ and the nonaxisymmetric
perturbation~(3.2).  The contour lines are drawn for
$\rho_*/\rho_{*~c}=10^{-0.3j}$ for $j=0,1,2,\cdots,10$ where
$\rho_{*~c}$ is 0.012, 0.15, and 0.99 for the three different times.  }
\end{figure}

\clearpage
\begin{figure}[t]
\epsfxsize=3in
\leavevmode
\epsffile{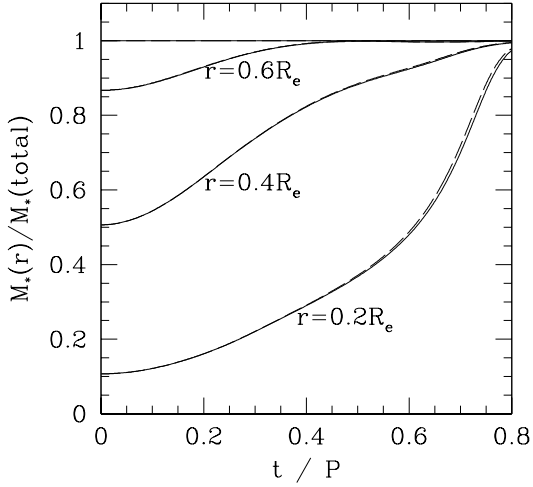}
\caption{Same as Fig. 7, but for 
star (A), $\Delta K/K=10\%$, with (solid line) and
without (dashed line) the nonaxisymmetric perturbation (3.2). } 
\end{figure}

\begin{figure}[t]
\epsfxsize=3in
\leavevmode
\epsffile{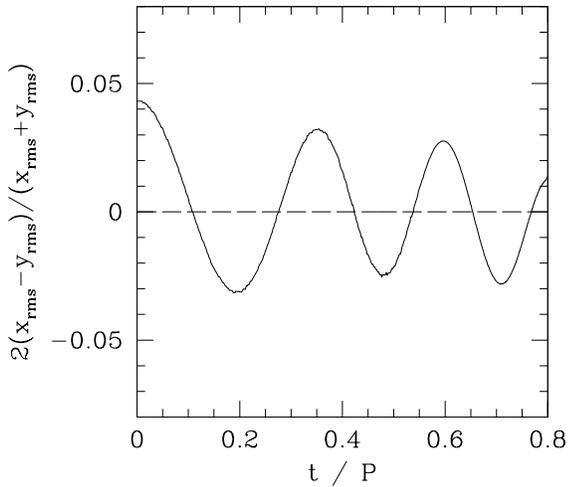}
\caption{Same as Fig. 11, but for the mean square axial length (see
Eq.~(3.3).  The solid and dashed lines denote simulations with and
without the nonaxisymmetric perturbation (3.2).}
\end{figure}

\begin{figure}[t]
\epsfxsize=3in
\leavevmode
\epsffile{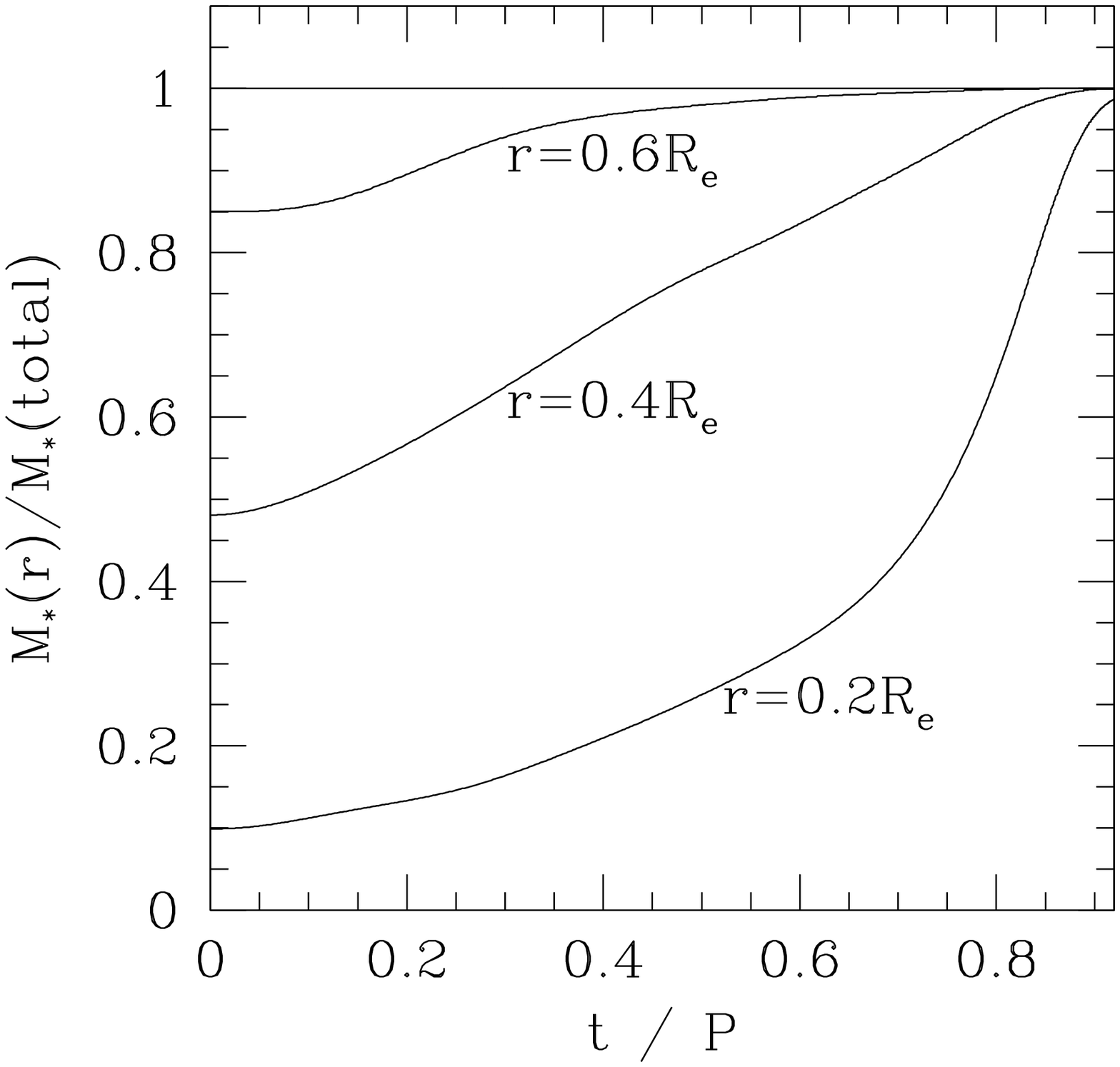}
\caption{Same as Fig. 7, but for collapse with the initial density
profile (3.4) (mimicing star (A) driven into instability by accretion 
of additional matter).}
\end{figure}


\begin{thebibliography}{99}
\bibitem{footrho} Throughout this paper, we use $\rho_c$ as the 
central rest mass density of equilibrium stars 
(initial data for our computation) and $\rho(r=0)$ as 
the central rest mass density of dynamically evolved stars (i.e., 
the latter changes during the evolution).
\bibitem{footnote} 
	Note that we recompute the constraint equations
	whenever we modify the initial equilibrium configurations
	(e.g. by depleting the pressure or by modifying the initial
	density profile).
\end{thebibliography}
\end{document}